\documentclass[aps,pra,amsmath,amssymb,floatfix,twocolumn, amsmath, superscriptaddress, twocolumn]{revtex4}
\usepackage{multirow}
\usepackage{bbold}
\usepackage{subfigure}
\usepackage{color}
\usepackage{mathrsfs}
\usepackage{hyperref}
\usepackage[normalem]{ulem}
\usepackage{bm}

\usepackage{amsfonts, relsize, color}
\usepackage{graphics}
\usepackage{graphicx}
\usepackage{subfigure}
\usepackage{hyperref}
\usepackage{color}

\begin{document}
\title{\bf Optical conductivity of an interacting Weyl liquid in the collisionless regime}
\author{Bitan Roy}
\affiliation{Department of Physics and Astronomy, Rice University, Houston, Texas 77005, USA}
\affiliation{Max-Planck-Institut f$\ddot{\mbox{u}}$r Physik komplexer Systeme, N$\ddot{\mbox{o}}$thnitzer Strasse 38, 01187 Dresden, Germany}

\author{Vladimir Juri\v ci\' c}
\affiliation{Nordita,   KTH Royal Institute of Technology and Stockholm University, Roslagstullsbacken 23,  10691 Stockholm,  Sweden}

\date{\today}
\begin{abstract}
Optical conductivity (OC) can serve as a measure of correlation effects in a wide range of condensed matter systems. We here show that the long-range tail of the Coulomb interaction yields a universal correction to the OC in a three-dimensional Weyl semimetal $\sigma(\Omega)=\sigma_0(\Omega)\left[ 1+\frac{1}{N+1} \right]$, where of $\sigma_0(\Omega)=Ne^2_0 \Omega/(12 h v)$ is the OC in the non-interacting system, with $v$ as the actual (renormalized) Fermi velocity of Weyl quasiparticles at frequency $\Omega$, and $e_0$ is the electron charge in vacuum. Such universal enhancement of OC, which depends only on the number of Weyl nodes near the Fermi level ($N$), is a remarkable consequence of an intriguing conspiracy among the quantum-critical nature of an interacting Weyl liquid, marginal irrelevance of the long-range Coulomb interaction and the violation of hyperscaling in three dimensions, and can directly be measured in recently discovered Weyl as well as Dirac materials. By contrast, a local density-density interaction produces a non-universal correction to the OC, stemming from the non-renormalizable nature of the corresponding interacting field theory.
\end{abstract}

\maketitle

\section{Introduction}

\begin{figure}[t!]
\subfigure[]{
\includegraphics[width=4cm,height=4.25cm]{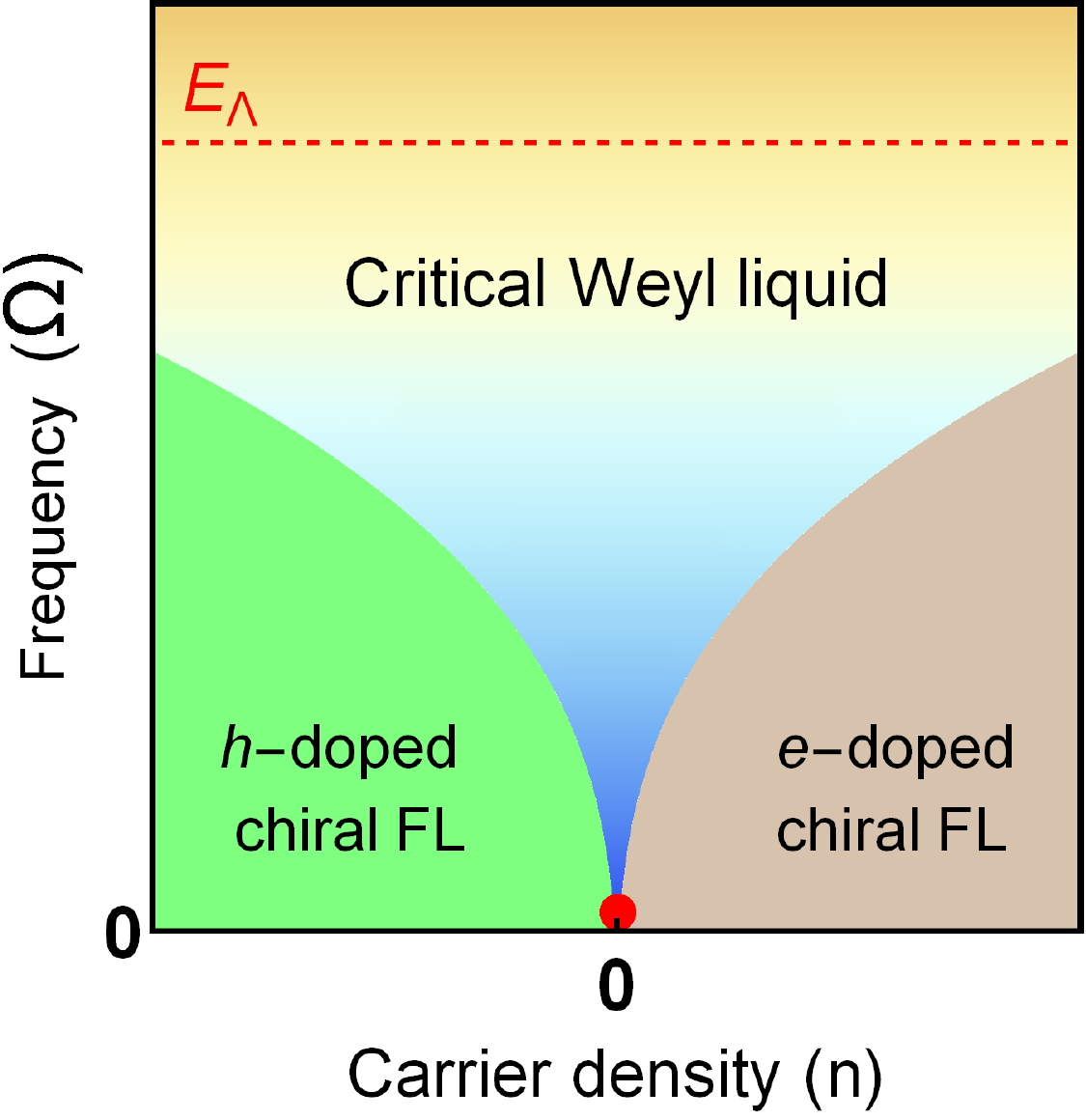}
\label{weylfan}
}
\subfigure[]{
\includegraphics[width=4cm,height=4.25cm]{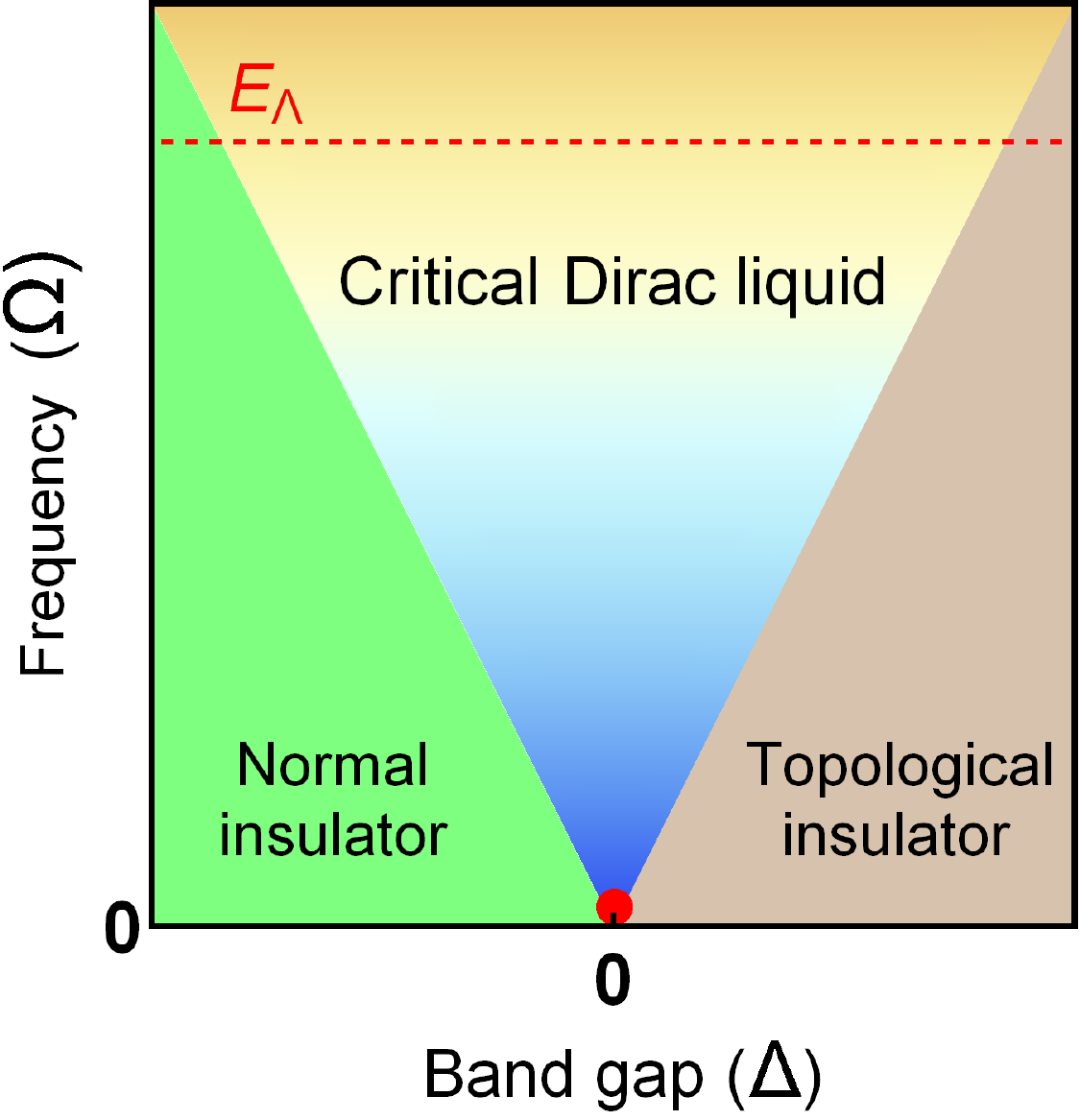}
\label{diracfan}
}
\caption{Quantum critical description of (a) Weyl and (b) Dirac liquid. Weyl [Dirac] semimetal can be represented as a quantum critical point (red dots) separating electron and hole doped chiral Fermi liquids (FLs)  [topological and normal insulators]. Signatures of Weyl or Dirac quasiparticles are present within the quantum critical fan (shaded regions), where the proposed universal scaling of optical conductivity with frequency [see Eq.~(\ref{OC:universal})] is operative. The crossover boundaries in (a) and (b) are respectively defined as $\Omega^\ast \sim v |n|^{1/3}$ and $|\Delta|$, up to interaction-driven corrections. At high frequencies ($\Omega \sim E_\Lambda$) imprints of Weyl or Dirac fermions gradually disappear and non-universal lattice details become important.
 }~\label{criticalfans}
\end{figure}

Optical conductivity (OC) stands as an indispensable experimental probe of electromagnetic response in a wide range of materials, including high-T$_c$ cuprate superconductors~\cite{basov2005}, heavy fermion compounds~\cite{Degiorgi1999,Haule2011}, Fe-based superconductors~\cite{Si2009,Degiorgi2011}, graphene~\cite{graphene:OC-1, graphene:OC-2, graphene:OC-3} and three-dimensional Weyl and Dirac systems~\cite{oc-exp-1,oc-exp-2,oc-exp-3,oc-exp-4,oc-exp-5,oc-exp-6}. This is so because charge dynamics has a direct impact on the OC, which then thus provides a rather comprehensive picture of electronic band structure, low-energy quasiparticle dynamics and nature of correlations in these systems.
In topological semimetals, which have recently attracted ample attention~\cite{TI:reivew-1,TI:reivew-2,weyl-review-1}, the imprint of electronic interactions on the OC may be important because undoped Weyl and Dirac semimetals at zero temperature ($T=0$) are inherently quantum critical states living in three dimensions [see Fig.~\ref{criticalfans}], where \emph{hyperscaling is violated}~\cite{zinn-justin}. Concomitantly, the thermodynamic potentials carry anomalous \emph{logarithmic corrections}~\cite{salam1975, blau1991, goswami-chakravarty, roy-sau, roy-goswami-sau, juricic-balatsky}. In addition, the long-range tail of the Coulomb interaction in these critical systems is marginally irrelevant, leading to a logarithmically slow vanishing of the fine structure constant due to the screening of the Coulomb charge and a simultaneous, also logarithmically slow growth of the Fermi velocity.

As we show, the Coulomb interaction causes a universal (independent of frequency and the fine structure constant) enhancement of the OC in an interacting Weyl liquid, arising from a subtle interplay between its marginal irrelevance and the violation of hyperscaling in three dimensions. The OC ($\sigma$) at frequency $\Omega$ is given by
\begin{equation}~\label{OC:universal}
\sigma(\Omega) = \sigma_0 (\Omega) \left[ 1+ \frac{1}{N+1}\right],
\end{equation}
after we account for the leading order correction due to the Coulomb interaction, and $\sigma_0(\Omega)\sim  \Omega/v$ is the OC in the noninteracting Weyl semimetal, featuring $N$ Weyl nodes in the Brillouin zone, with $v$ as the renormalized (experimentally measured) Fermi velocity of the Weyl quasiparticles at frequency $\Omega$ in the interacting system. This is the central result of our work. 

Although Coulomb interaction enhances the OC, for sufficiently large number of Weyl nodes ($N \gg 1$), the interaction driven correction to the OC scales as $\sim 1/N$, which then vanishes as $N \to \infty$. Such peculiar scaling stems from the dynamic screening of the electronic charge by massless Weyl fermions in the medium. Hence, the scaling in Eq.~(\ref{OC:universal}) can be viewed as the leading term of a systematic and controlled $1/N$-expansion of the OC in an interacting Weyl semimetal. Since the long-range Coulomb interaction is expected to be always marginally irrelevant [see, for example, Ref.~\cite{prokofev} for such conclusion in two dimensions], we are compelled to believe that interaction mediated enhancement of OC possibly remains valid beyond the leading order in $1/N$ and thus should be observable in recently discovered Weyl and Dirac materials~\cite{vishwanath, dai-zrte, taas-2, nbas-1, felser-1, nbp-1, tap-1, cdas, nabi, goswami-roy-dassarma}. Recent experiment on ZrTe$_5$~\cite{oc-exp-3}, a predicted Dirac semimetal~\cite{dai-zrte}, found a large enhancement of the OC. By contrast, a weak short range interaction is an irrelevant perturbation at the Weyl or Dirac quantum critical point (QCP), see Fig.~\ref{criticalfans}, and provides only a non-universal correction to the OC which rapidly vanishes as $\Omega \to 0$ [see Eq.~(\ref{eq:OC-correction})].

In the next section we introduce the low-energy theory of an interacting Weyl liquid, and discuss the scaling of OC in non-interacting system. In this section, we also briefly review the renormalization group flow of Coulomb interaction. Sec.~\ref{opticalconductivity} is devoted to the discussion on the scaling of OC and its correction due to Coulomb interaction to the leading order. We summarize our findings and present discussion on related systems (such as Dirac semimetals) in Sec.~\ref{discussion}. Technical details of our analysis are displayed in the Appendices~\ref{RG:leadingorder}-\ref{append:dielectric}.

\section{Interacting Weyl fermions and optical conductivity}~\label{Weylfermions}

Weyl semimetal can be envisioned as the simplest example of a QCP, separating an electron- and a hole-doped chiral Fermi liquids, that supports linearly dispersing sharp low-energy quasiparticles, with dispersion $E_{\bf k}=v |{\bf k}|$, up to a high energy cutoff $E_\Lambda$ [see Fig.~\ref{weylfan}]. The Weyl QCP is therefore characterized by the dynamical exponent $z=1$, which determines the relative scaling between energy and momentum. The corresponding Euclidean action is
\begin{equation}
S_0= \int d\tau d{\bf r}\,\psi^\dagger(\tau, {\bf r}) \,[\partial_\tau \pm (-i) v {\bm \sigma} \cdot {{\bm \nabla}} + \mu]\, \psi(\tau, {\bf r}),
\end{equation}
with $\tau$ as the imaginary time and $\pm$ denoting the two chiralities of the Weyl cones which on a lattice always appear in pairs~\cite{nielsen}. Here, $\mu$ is the chemical potential, measured from the apex of the conical dispersion, ${\bm \sigma}$s are standard Pauli matrices acting on the two-component spinors $\psi(\tau, {\bf r})$ representing (pseudo-)spin. The chemical potential with positive scaling dimension $[\mu]=z=1$ is the relevant perturbation at Weyl QCP point that controls a quantum phase transition (QPT), characterized by the correlation length exponent $\nu=1$, from a hole- to an electron-doped chiral Fermi liquid. Together, these two exponents ($\nu$ and $z$) define the universality class of this QPT, as well as determine the crossover boundaries at frequency ${\Omega^\ast}\sim v|n|^{1/3}$ or temperature $T^\ast \sim(\hbar v/k_B)|n|^{1/3}$, among various phases in terms of the carrier density, $n$; see Fig.~\ref{weylfan}~\cite{Schmalian}. The signature of Weyl fermions in transport and thermodynamic quantities can therefore be observed for $\Omega>\Omega^\ast$ and $T> T^\ast$. Specifically, we here focus on the OC of such critical Weyl liquid in the collisionless regime ($\Omega \gg T$), with $T=0$ from outset.

The scaling form of the OC ($\sigma$) can be inferred from the gauge invariance which dictates that  $[\sigma]=d-2$ exactly \cite{Sachdev-book}, or $\sigma(\Omega)=\sigma_Q {\ell}^{-1}$ in units of quantum conductance $\sigma_Q=e_0^2/h$. Here, $\ell$ is a characteristic length scale inside the Weyl critical fan [shaded region in Fig.~\ref{weylfan}] at finite frequency, and thus $\ell \sim v/\Omega$. The OC of a noninteracting Weyl liquid is then given by $\sigma_0(\Omega)= \sigma_Q c_0 \Omega/v$, with $c_0=N/12$ as a universal number~\cite{goswami-chakravarty, hosur, rosenstein, roy-juricic-dassarma} and the system behaves as a \emph{power-law insulator}, since $\sigma_0(\Omega \to 0) \to 0$ [see Appendix~\ref{OC:noninteracting}].

In the presence of generic density-density interaction, captured by the imaginary-time action
\begin{equation}
S_{\rm int}=\int d\tau d{\bf r} d{\bf r}' \rho(\tau,{\bf r})V({\bf r}-{\bf r}')\rho(\tau,{\bf r}'),
\end{equation}
where $\rho(\tau,{\bf r})=\psi^\dagger(\tau,{\bf r})\psi(\tau,{\bf r})$ is the electronic density, the correction to the OC depends crucially on its range. For the long-range Coulomb interaction $V({\bf r}-{\bf r}')=e^2/|{\bf r}-{\bf r}'|$, and thus $[e^2]=z-1$, implying that the dimensionless coupling is the fine structure constant $\alpha=  e^2/v$. Furthermore, in the reciprocal space the Coulomb interaction $V({\bf k})\sim e^2/k^2$ is an analytic function of the momentum and therefore charge is dynamically screened by Weyl fermions, as opposed to the situation in two dimensions~\cite{Gonzalez1994}, which together with the logarithmically slow increase of the Fermi velocity makes the fine structure constant marginally irrelevant in a Weyl fluid. 

These key features, even though believed to be true in general, can qualitatively be appreciated from the leading order flow equations for $v$ and $\alpha$ respectively given by~\cite{goswami-chakravarty, hosur, rosenstein, throckmorton} [see Appendix~\ref{RG:leadingorder}]
\begin{equation}~\label{RG:leadingorder}
\frac{dv}{dl}= \frac{\alpha v}{3\pi}, \: \: \frac{d \alpha}{dl}=-\frac{N+1}{3 \pi} \alpha^2,
\end{equation}
where $l=\log(E_\Lambda/\Omega)$ is the logarithm of the renormalization group length scale. On the other hand, for a contact interaction $V({\bf r}-{\bf r}')=g_0\delta({\bf r}-{\bf r}')$, the scaling dimension of the coupling is $[g_0]=z-D$~\cite{roy-goswami-sau, roy-dassarma, nandkishore, roy-goswami-juricic}, which makes it irrelevant close to the Weyl QCP in $D=3$. Thus, the dimensionless short range coupling $g=g_0\Omega^2/v^3$, yielding $dg/dl=-2 g$ to the leading order. Consequently, while its long-range tail provides the leading correction to physical observables in the noninteracting system, such as the OC as we demonstrate here, the short-range pieces of the Coulomb interaction give rise to only subleading corrections.

\section{Scaling and correction to optical conductivity}~\label{opticalconductivity}

General scaling arguments suggest that OC in an interacting Weyl liquid assumes the following form in terms of the renormalized couplings and Fermi velocity
\begin{equation}~\label{scaling:OC}
\sigma(\Omega,\alpha,g)=\sigma_0(\Omega)F(\alpha,g),
\end{equation}
where $F(g,\alpha)$ is a universal scaling function, with $F(0,0)=1$. We then recover the OC in the noninteracting system. Since in three spatial dimensions hyperscaling hypothesis is violated, the above scaling function receives logarithmic corrections (besides the usual power-law ones), which to the order $n$ in the perturbation theory have the form
\begin{equation}\label{eq:scaling-function1}
F_n(g,\alpha)=\sum_{m=0}^n \left(\alpha^n C_{n,m}+g^n G_{n,m} \right) \; \log^m\left(\frac{E_\Lambda}{\Omega}\right),
\end{equation}
with $C_{n,m}$ and $G_{n,m}$ as the real coefficients, and $F(g,\alpha)=\sum_{n\geq0}F_n(g,\alpha)$. We here determine these coefficients perturbatively to the leading order in the coupling constants ($\alpha$ and $g$), i.e. for $n=1$.

To find the OC in the presence of interactions, we first compute the correction to the current-current correlation function
$\Pi_{\mu\nu}(i\Omega,{\bf q})$, with $\Omega$ as Matsubara frequency, ${\bf q}$ as momentum,
which is the Fourier transform of $\Pi_{\mu\nu}(\tau,{\bf r})=\langle j_\mu(\tau,{\bf r}) j_\nu(0,0) \rangle$, and $\mu,\nu=0,1,2,3$. Here, ``four"-current $j_\mu(\tau,{\bf r})=[\rho(\tau,{\bf r}),{\bf j}(\tau,{\bf r})]$, with the spatial components ${\bf j}(\tau,{\bf r})=v \psi^\dagger(\tau,{\bf r}){\bm \sigma}\psi(\tau,{\bf r})$. The charge conservation
$-i\partial_\tau\rho+{\bm\nabla}\cdot{\bf j}=0$
then implies
\begin{equation}~\label{chargeconservation}
-(i\Omega)^2\Pi_{00}(i\Omega,{\bf q})+q_l q_m \Pi_{lm}(i\Omega,{\bf q})=0,
\end{equation}
with $l,m=1,2,3$, which constraints physically relevant regularizations of the theory. We here employ the dimensional regularization scheme in spatial dimensions $D=3-\epsilon$, as it manifestly preserves the $U(1)$ symmetry of the theory~\cite{Hooft, peskin} and obtain
\begin{equation}\label{eq:polarization1}
\Pi_{00}(i\Omega,{\bf q})=q^2{\tilde \Pi}(i\Omega),\,\,\Pi_{lm}(i\Omega,0)=-\delta_{lm}\Omega^2{\tilde \Pi}(i\Omega),
\end{equation}
with
\begin{equation}\label{eq:polarization2}
{\tilde \Pi}(i\Omega)=\frac{N}{12\pi^2v}\left[ \frac{e^2}{6\pi v}\left(\frac{1}{\epsilon^2}+\frac{b}{\epsilon}\right) + \frac{g_0\Omega^2}{24\pi^2v^3} \left(\frac{1}{\epsilon^2}+\frac{a}{\epsilon}\right) \right],
\end{equation}
where $a=\left[5-3\gamma_E+3\log(4\pi)\right]/3\approx 3.62069$, $b=-\left[1+2\gamma_E-2\log(4\pi)\right]/2\approx 1.454$, and $\gamma_E\approx0.577$ is the Euler-Mascheroni constant [see Appendix~\ref{OC:short-range}, ~\ref{OC:long-range}, ~\ref{OC:alternative}]. The terms proportional to $1/\epsilon \sim \log(E_\Lambda /\Omega)$ capture the logarithmic divergent pieces of the current-current correlator. This result is consistent with the charge conservation condition, displayed in Eq.~(\ref{chargeconservation}). Subsequently, we use the Kubo formula
\begin{equation}\label{eq:Kubo}
\sigma_{lm}(\Omega)=2\pi\sigma_Q\lim_{\delta\rightarrow0}
\frac{\Im\Pi_{lm}(i\Omega\rightarrow\Omega+i\delta)}{\Omega}
\end{equation}
to find interaction driven leading order correction to the OC, as given by Eq.~(\ref{eq:scaling-function1}), with [see Appendix~\ref{OC:short-range}, ~\ref{OC:long-range}, ~\ref{OC:alternative}]
\begin{equation}~\label{coefficients}
C_{1,1}=\frac{1}{3\pi}, C_{1,0}=-\frac{b}{6\pi},  G_{1,1}=-\frac{1}{12\pi^2}, G_{1,0}=\frac{a}{24\pi^2}.
\end{equation}
We note that computation of the current-current correlator using the hard cut-off method violates the charge conservation condition~\cite{herbut-juricic-vafek-2}, and may lead to unreliable values of the above coefficients~\cite{rosenstein}.

Finally, we recall that the couplings entering the scaling function [see Eq.~(\ref{eq:scaling-function1})] are the renormalized ones and thus scale dependent. From the leading order renormalization group flow equations [see Appendix~(\ref{RG:leadingorder})], we obtain
\begin{eqnarray}
\alpha(\Omega) \approx \frac{3 \pi}{(N+1) \log\left(\frac{E_\Lambda}{\Omega}\right)}, \:\:
g(\Omega)=\hat{g}_0 \left(\frac{\Omega}{E_\Lambda}\right)^2,
\end{eqnarray}
where $\hat{g}_0$ is the dimensionless bare short-range coupling. Then together with Eqs.~(\ref{eq:polarization1}),~(\ref{eq:polarization2}),~(\ref{eq:Kubo}), above running couplings in turn yield the leading correction to the OC in a Weyl liquid, giving
\begin{align}\label{eq:OC-correction}
\sigma(\Omega)&=\sigma_0(\Omega)\left[1+\frac{1}{N+1}-\frac{b}{2 (N+1) \log\left(\frac{E_\Lambda}{\Omega}\right)}\right.\nonumber\\
&\left.-\frac{\hat{g}_0 \; \Omega^2}{12\pi^2 E^2_\Lambda}
\left\{ \log\left(\frac{E_\Lambda}{\Omega}\right)-\frac{a}{2}\right\} \right],
\end{align}
which simplifies to Eq.~(\ref{OC:universal}) for $\Omega \ll E_\Lambda$.

Therefore, the long-range tail of the Coulomb interaction yields a universal (independent of frequency and the strength of the fine structure constant) correction to the OC of the noninteracting Weyl fluid, which is a remarkable consequence of an intriguing conspiracy among the quantum-critical nature of a Weyl semimetal [see Fig.~\ref{weylfan}], marginal irrelevance of the long-range Coulomb interaction [see Eq.~(\ref{RG:leadingorder})] and the violation of hyperscaling in three dimensions [see Eqs.~(\ref{scaling:OC}),~(\ref{eq:scaling-function1}) and (\ref{coefficients})]. In particular, the logarithmically slow decrease of the fine structure constant at low energy precisely cancels the perturbatively obtained logarithmic correction to the OC, ultimately producing a finite result. This outcome is staunchly suggestive of the renormalizability of the field theory describing an interacting Weyl liquid in the presence of only long-range Coulomb interaction. Note that such correction is operative in the entire quantum-critical regime of the Weyl fluid, shown in Fig.~\ref{weylfan} [the shaded regime], making our prediction relevant for real Weyl materials where chemical potential often (if not always) is placed away from the band touching diabolic points. For an analogous problem in two dimensions, the fine structure constant is also marginally irrelevant, but hyperscaling holds, giving rise to a positive, but logarithmically slowly vanishing correction to the OC~\cite{herbut-juricic-vafek-1, herbut-juricic-vafek-2}. 

On the other hand, the short-range piece of the Coulomb interaction in a Weyl fluid produces only a power-law correction of the conductivity, which ultimately vanishes in the $\Omega \rightarrow 0$ limit. Moreover, the explicit dependence of this correction on the ultraviolet cutoff ($E_\Lambda$) cannot be eliminated through a redefinition of the bare coupling constant, reflecting non-renormalizability of the field theory of (quasi-)relativistic fermions coupled via short-range interaction in three dimensions~\cite{zinn-justin, peskin}. In other words, the lattice details, such as the bare strength of the coupling and the ultraviolet cutoff ($E_\Lambda$), enter this contribution to the conductivity, which makes the correction to the OC due to local interaction a nonuniversal quantity. Otherwise, the short-range piece causes a reduction in the OC, which is qualitatively consistent with the fact that it can trigger a QPT from a Weyl semimetal to a translational-symmetry breaking  axionic insulator, when sufficiently strong~\cite{roy-dassarma, nandkishore, roy-goswami-juricic}.

The Coulomb correction to OC scales as $\sim 1/N$ for $N \gg 1$ and therefore vanishes in the limit when the Brillouin zone accommodates a large number of Weyl points ($N \to \infty$), since in this limit the Coulomb interaction suffers complete dynamic screening by massless Weyl fermions. Furthermore, using the general scaling argument we can speculate that the $n^{\rm th}$ order correction goes as $\sim 1/N^n$, with the coefficients that remain to be determined for $n>1$. Nevertheless, existence of a plethora of Weyl compounds with diverse flavor number ($N$), such as $N=24$ and $ N=6$ respectively in all-in all-out and spin-ice ordered Weyl phase in 227 pyrochlore iridates~\cite{vishwanath, goswami-roy-dassarma}, $N=24$ in inversion asymmetric Weyl semimetal (such as TaAs, NbAs, etc.)~\cite{taas-2, nbas-1, felser-1, nbp-1, tap-1, cdas, nabi}, and topological Dirac semimetal (that at low energies can be considered as two superimposed Weyl semimetals) such as Cd$_2$As$_3$ and Na$_3$Bi with $N=4$~\cite{cdas,nabi}, endows an unprecedented opportunity to extract the scaling of OC as a function of $N$ and test the validity of our prediction [Eq.~(\ref{OC:universal})], with prior notion of the Fermi velocity ($v$) available from the APRES measurements, for example.

\section{Discussions}~\label{discussion}

For completeness, we also report the correction to the dielectric constant in Weyl liquid, given by 
\begin{equation}~\label{dielectric}
\varepsilon(\Omega)=1+\frac{2N e_0^2}{3 h v} \left[1+\frac{1}{N+1} \right] \; \log\left( \frac{E_\Lambda}{\Omega}\right),
\end{equation}
due to the long-range Coulomb interaction, which can directly be obtained from Eq.~(\ref{OC:universal}) by applying the Kramers-Kronig relation [see Appendix~\ref{append:dielectric}]. Notice that a logarithmic enhancement of $\varepsilon(\Omega)$, observed in a recent experiment~\cite{oc-exp-4}, is a clear manifestation of the violation of the hyperscaling hypothesis in three dimensions.

Our conclusions regarding the scaling of OC and dielectric constant are also directly applicable for interacting massless Dirac fermions that can be represented as a band gap ($\Delta$) tuned QCP, separating two topologically distinct insulating phases in three dimensions, see Fig.~\ref{diracfan}. Since Dirac semimetal supports linearly dispersing quasiparticles and the scaling dimension of the band gap in this system is $[\Delta]=1$, the Dirac QCP is also characterized by $z=1$ and $\nu=1$.  Consequently, in the presence of Coulomb interaction the correction to the OC is given by Eq.~(\ref{OC:universal}) or ~(\ref{eq:OC-correction}), and that for the dielectric constant takes the form of Eq.~(\ref{dielectric}), but with a modification $N \to 2 N$, and $N$ now counts the number of four-component Dirac fermions. Therefore, $N=1$ for Bi$_2$Se$_3$, Bi$_{1-x}$Sb$_x$, Hg$_{1-x}$Cd$_x$Te, ZrTe$_5$, $N=4$ for Pb$_{1-x}$Sn$_x$Te~\cite{TI:reivew-1, TI:reivew-2,dornhaus}, and  $N=3$ for SmB$_6$, YbB$_6$~\cite{roy-dzero}. Notice that even on the insulating sides of the phase diagram ($|\Delta| \neq 0$) and/or when the chemical potential lies in the valence or conduction band, the proposed scaling of OC remains valid as long as $\Omega>\Omega^\ast \sim |\Delta|$ or $v|n|^{1/3}$, see Fig.~\ref{criticalfans}. Therefore, existence of ample Dirac materials with different number of Dirac nodes ($N$) and residing in the close proximity to the topological QPT, which can also be tuned externally by applying hydrostatic pressure or changing the chemical composition, constitutes an ideal platform to test the validity of proposed scaling of the OC and dielectric constant.

To summarize, we here explicitly demonstrate that an intriguing confluence among the quantum critical nature of an interacting Weyl or Dirac liquid, hyperscaling violation in three dimensions and marginal nature of the long-range Coulomb interaction endows these systems with a universal  interaction mediated enhancement of the OC. The scaling of this correction with the flavor number can directly be probed in a large number of discovered and predicted Dirac and Weyl materials~\cite{TI:reivew-1, TI:reivew-2, weyl-review-1, vishwanath, dai-zrte, taas-2, nbas-1, felser-1, nbp-1, tap-1, cdas, nabi, goswami-roy-dassarma} as well as in numerical simulations~\cite{katsnelson}, making our results relevant to recent and ongoing experiments ~\cite{oc-exp-1,oc-exp-2,oc-exp-3,oc-exp-4,oc-exp-5,oc-exp-6}. Finally, our findings may also motivate future investigations of the interaction effects on magneto-transport and on hydrodynamic transport in interacting Weyl/Dirac liquids.

\acknowledgements

B. R. was supported by the Welch Foundation Grant No.~C-1809 and by NSF CAREER Grant no.~DMR-1552327 of Matthew S. Foster (Rice University).

\appendix

\section{Leading order renormalization group flow equations for Fermi velocity and fine-structure constant}~\label{RG:leadingorder}

The Euclidean action for the Weyl quasiparticles interacting with the long range Coulomb interaction has the form
\begin{eqnarray}\label{eq:action-Weyl-Coulomb}
S &=& \int d\tau d{\bf r}\psi^\dagger(\tau,{\bf r})[\partial_\tau-i a_0-i v {\bm \sigma}\cdot{\bm\nabla}]\psi(\tau,{\bf r}) \nonumber \\
&+&\frac{1}{2}\int d{\bf r} a_0({\bf r})\frac{|\nabla|^2}{2\pi e^2} a_0({\bf r}),
\end{eqnarray}
where $\tau$ is the imaginary time, ${\bf r}$ is the spatial coordinate, $\psi(\tau,{\bf r})$ is a two-component Weyl spinor, $v$ is the Fermi velocity of the Weyl quasiparticles, and ${\bm \sigma}$s are the Pauli matrices. The partition function is $\mathcal{Z}=\int \mathcal{D}\Phi\, e^{-S[\Phi]}$, where $\Phi$ denotes all the fields in the action.
Auxiliary gauge field $a_0$ is chosen so that after integrating it out, the Coulomb interaction has the usual $1/k^2$ form in three spatial dimensions
\begin{eqnarray}
S_c &=&\int d^3{\bf k}\; d^3{\bf k'}\;\rho({\bf k})\;\frac{2\pi e^2}{|{\bf k}-{\bf k}'|^2}\;\rho({\bf k}') \nonumber \\
&\equiv& \int d{\bf k}\; d{\bf k'}\;\rho({\bf k})\;V_C({\bf k}-{\bf k}')\;\rho({\bf k}'),
\end{eqnarray}
with $\rho({\bf k})$ as the density operator in momentum space, and $V_C({\bf k})=2\pi e^2/{\bf k}^2$. 
The propagators for the Weyl fermion and the gauge field in terms of a Matsubara frequency $i\omega$ and a momentum ${\bf k}$, respectively, read
\begin{equation}
G_f(i\omega,{\bf k})=\frac{i\omega+v{\bm \sigma}\cdot{\bf k}}{\omega^2+v^2k^2},
\end{equation}
\begin{equation}\label{eq:gauge-propagator}
G_{a_0}({\bf k})=\frac{2\pi e^2}{{\bf k}^2}.
\end{equation}

We now perform a renormalization group analysis using the dimensional regularization in $D=3-\epsilon$ dimensions and a minimal subtraction scheme to find the leading order flow equation for the Fermi velocity and the Coulomb charge.
The one-loop $\beta$ function for the Fermi velocity follows from the leading order correction to the self-energy for Weyl fermions due to the long-range tail of the Coulomb interaction which reads as
\begin{equation}
\Sigma\left( i\omega, {\mathbf q} \right) = i^2 \int \frac{d^D {\mathbf k}}{(2 \pi)^D} \int^{\infty}_{-\infty} \frac{d\omega}{2 \pi} V_C({\bf k})G_f(i(\omega+\Omega),{\bf k}+{\bf q}).
\end{equation}
An explicit calculation then yields
\begin{eqnarray}\label{eq:self-energy}
&&\Sigma\left( i\omega, {\mathbf q} \right) \nonumber \\
&=& -\int \frac{d^D {\mathbf k}}{(2 \pi)^D} \int^{\infty}_{-\infty} \frac{d\omega}{2 \pi} \frac{2 \pi e^2}{|{\mathbf k}|^2} \; \frac{i (\omega+\Omega) + v {\boldsymbol \sigma} \cdot ({\mathbf k}+{\mathbf q}) }{(\omega+\Omega)^2+ v^2 ({\mathbf k}+{\mathbf q})^2} \nonumber \\
&=&-\frac{1}{2 v} \int \frac{d^D {\mathbf k}}{(2 \pi)^D} \frac{2 \pi e^2}{|{\mathbf k}|^2} \; \frac{{\boldsymbol \sigma} \cdot ({\mathbf k}+{\mathbf q})}{|{\mathbf k}+{\mathbf q}|} \nonumber \\
&=& -\frac{\pi e^2}{2 } \int^1_0 dx \; \int \frac{d^D {\mathbf k}}{(2 \pi)^D} \; \frac{x^{-1/2} (1-x)  \; {\boldsymbol \sigma} \cdot {\mathbf q}}{\left[ k^2 + x(1-x) q^2 \right]^{3/2}} \nonumber \\
&=&-\frac{{\boldsymbol \sigma} \cdot {\mathbf q}}{|{\mathbf q}|^{3-D}} \; \frac{\pi e^2 \Gamma\left( \frac{3-D}{2}\right)}{2  (4\pi)^{D/2} \Gamma\left(\frac{3}{2}\right) }  \int^1_0 dx \left(1-x\right)^{\frac{D-1}{2}} x^{\frac{D-4}{2}}\nonumber\\
&=&-\frac{{\boldsymbol \sigma} \cdot {\mathbf q}}{|{\mathbf q}|^{3-D}}\;\; \frac{\pi e^2 \Gamma\left( \frac{3-D}{2}\right)}{2  (4\pi)^{D/2} \Gamma\left(\frac{3}{2}\right) }\;\frac{\Gamma\left(\frac{D}{2}-1\right)\Gamma\left(\frac{D+1}{2}\right)}{\Gamma\left(D-\frac{1}{2}\right)} \nonumber \\
&=& -\left( \frac{e^2}{3  \pi} \right) \: \left[ \frac{{\boldsymbol \sigma} \cdot {\mathbf q}}{|{\mathbf q}|^{\epsilon}} \right] \: \frac{1}{\epsilon},
\end{eqnarray}
for $D=3-\epsilon$. On the other hand, the one-loop polarization bubble, leading to charge renormalization, reads as
\begin{eqnarray}
\Pi\left( i\omega, {\mathbf q}\right) &=& - i^2 {\rm \bf Tr}\int \frac{d^D {\mathbf k}}{(2 \pi)^D} \int^{\infty}_{-\infty} \frac{d\omega}{2 \pi}
\big[ \sigma_0 \; G_f(i\omega,{\bf k}) \nonumber \\
&\times& \sigma_0\; G_f(i(\omega+\Omega),{\bf k}+{\bf q}) \big].
\end{eqnarray}
To find the renormalization factor that ultimately gives the beta function for the charge, it is sufficient to keep the terms in the polarization up to $q^2$ order, leading to
\begin{eqnarray}\label{eq:bubble}
\Pi\left( i\omega, {\mathbf q}\right) &=& - i^2 \;  {\rm \bf Tr} \int \frac{d^D {\mathbf k}}{(2 \pi)^D} \int^{\infty}_{-\infty} \frac{d\omega}{2 \pi} \; \bigg[ \sigma_0 \; \frac{i \omega + v {\boldsymbol \sigma} \cdot {\mathbf k} }{\omega^2+ v^2 {\mathbf k}^2}  \nonumber \\
&\times& \sigma_0 \frac{i (\omega+\Omega) + v {\boldsymbol \sigma} \cdot ({\mathbf k}+{\mathbf q}) }{(\omega+\Omega)^2+ v^2 ({\mathbf k}+{\mathbf q})^2} \bigg] \nonumber \\
&=& -\;\Pi_{00}(i\omega,{\bf q})=-\frac{N q^2}{12 \pi^2 v} \left[ \frac{q^{-\epsilon}}{\epsilon }+ \mathcal{O}(1)  \right],
\end{eqnarray}
where the density-density correlator $\Pi_{00}(i\omega, {\bf q})$ to the $q^2$ order is given by Eq. (\ref{eq:polarization-bubble-cond-1}).

From Eq.~(\ref{eq:self-energy}), after re-exponentianting it, introducing the wavefunction renormalization $Z_\psi$ and the renormalization factor for the velocity $Z_v$ via $Z_v v=v_0$, with $v_0$ as the bare velocity, we find
\begin{equation}
Z_\Psi(-i\omega+Z_v v {\bm \sigma}\cdot {\bf q})+\frac{e^2 q^{-\epsilon}}{3\pi v \epsilon}v {\bm \sigma}\cdot {\bf q}=-i\omega +v {\bm \sigma}\cdot {\bf q},
\end{equation}
which then yields
\begin{equation}
Z_\psi=1,\,\, Z_v=1-\frac{e^2}{3\pi v \epsilon}q^{-\epsilon}.
\end{equation}
Using renormalization condiction $Z_v v=v_0$, we then readily obtain the infrared $\beta-$function for the velocity to the leading order
\begin{equation}\label{eq:beta-velocity}
\beta_v\equiv-\frac{dv}{d\log q}=\frac{e^2}{3\pi}=\frac{v \alpha}{3 \pi},
\end{equation}
where $\alpha=e^2/v$ is the fine structure constant.

The flow equation for charge is obtained from Eq.\ (\ref{eq:bubble}) after re-exponentiating it, and recalling that the form of the action for the gauge field is given by Eq. (\ref{eq:action-Weyl-Coulomb}). The renormalization condition for the charge then reads
\begin{equation}
\frac{1}{2\pi e_0^2}+\frac{N q^{-\epsilon}}{6\pi^2 v\epsilon}=\frac{1}{2\pi e^2},
\end{equation}
with $e_0$ denoting the bare charge, from which we find  renormalization constant for the charge $Z_{e^2}e^2=e_0^2$ to be of the form
\begin{equation}
Z_{e^2}=1+\frac{N e^2}{3\pi v\epsilon}q^{-\epsilon},
\end{equation}
which yields the leading order beta function for the charge
\begin{equation}\label{eq:charge-beta}
\beta_{e^2}=-\frac{d e^2}{d \log k}=-\frac{N e^4}{3\pi v}.
\end{equation}
Finally, using Eqs. (\ref{eq:beta-velocity}) and (\ref{eq:charge-beta}), we obtain the flow equation for the
fine structure constant $\alpha=e^2/v$
\begin{equation}~\label{beta:alpha}
\beta_\alpha=-\frac{1}{3\pi}(N+1)\alpha^2\equiv -A\alpha^2,
\end{equation}
which therefore yields  the coefficient $A=(N+1)/3\pi$ that we use in the main text.
\\

On the other hand, the imaginary time action in the presence of only local density-density interaction reads as
\begin{equation}
S_{SR}=g_0 \int d\tau \; d{\bf r} \: \left[ \psi^\dagger(\tau,{\bf r}) \psi(\tau,{\bf r}) \right]^2,
\end{equation}
where $g_0$ denotes the strength of contact interaction. In this notation $g_0>0$ represents repulsive interaction. The scaling dimension of any short-range interaction is $[g_0]=z-D$, implying $[g_0]=-2$ (and thus an irrelevant perturbation) in a three-dimensional ($D=3$) Weyl liquid ($z=1$), yielding the following leading order beta function (infrared) for the dimensionless coupling constant, defined as $g=g_0 \Omega^2/v^3$ for example,
\begin{equation}~\label{beta:g0}
\beta_{g}=-2 \; g + {\mathcal O}(g^2).
\end{equation}



\onecolumngrid

\section{Optical conductivity in noninteracting system}~\label{OC:noninteracting}

 In two subsequent sections, we will first provide the detailed derivation of the components polarization tensor corresponding to (a) current-current correlator at zero momentum and (b) density-density correlator at finite but small momenta, which enter the Kubo formula for the optical conducitivity. Subsequently, we present a proof that the polarization function in a noninteracting system computed using dimensional regularization about $D=3$ spatial dimensions is consistent with the charge conservation.   Finally, we show how to obtain the expression for OC from the expression of the current-current correlator at zero momentum and finite frequency after the analytic continuation.

\subsection{Current-current correlator at ${\bf q}=0$}

The current-current correlation function for noninteracting Weyl fermions at a finite Matsubara frequency and a momentum  reads
\begin{equation}\label{eq:current correlator-1}
\Pi_{lm}\left(i\Omega,{\bf q} \right) =- \int \frac{d^D {\mathbf k}}{(2\pi)^D} \; \int^{\infty}_{-\infty} \frac{d\omega}{2 \pi} \; { {\rm \bf Tr}} \;\left[\left(v \sigma_l\right) \; G_f(i(\omega+\Omega),{\bf k}+{\bf q})\;\left( v \sigma_m \right)\;  G_f(i\omega,{\bf k})\right],
\end{equation}
since the current operator is of the form  ${\bf j}=\psi^\dagger {v \bm \sigma}\psi$.
To find optical conductivity from the current-current correlator we compute it only at the zero momentum, which yields
\begin{eqnarray}\label{eq:current-correlator-free}
\Pi_{lm}\left(i\Omega,0 \right) &=&- \int \frac{d^D {\mathbf k}}{(2\pi)^D} \; \int^{\infty}_{-\infty} \frac{d\omega}{2 \pi} \; {\bf Tr} \; \left[ \left( v \sigma_l \right) \; \frac{i (\omega+\Omega) + v {\boldsymbol \sigma} \cdot {\mathbf k}}{(\omega+\Omega)^2 + v^2 {\mathbf k}^2 } \; \left( v \sigma_m \right) \; \frac{i \omega + v {\boldsymbol \sigma} \cdot {\mathbf k}}{\omega^2 + v^2 {\mathbf k}^2 }  \right] \nonumber \\
&=& -2 N \delta_{l,m} \; v^2 \; \int \frac{d^D {\mathbf k}}{(2\pi)^D} \; \int^{\infty}_{-\infty} \frac{d\omega}{2 \pi} \;  \left[ \frac{-\omega(\omega+\Omega) + \left( \frac{2}{D}-1 \right) v^2 {\mathbf k}^2}{\left[ (\omega+\Omega)^2 + v^2 {\mathbf k}^2 \right] \; \left[ \omega^2 + v^2 {\mathbf k}^2  \right]} \right] \nonumber \\
&=& -\frac{4 N}{v^{D-2}} \delta_{l,m} \; \left( \frac{1}{D}-1 \right) \; \frac{2 \pi^{D/2}}{\Gamma\left( \frac{D}{2}\right) (2 \pi)^D} \; \int^{\infty}_0 \; dk \: \frac{ k^D}{4 k^2+\Omega^2} \nonumber \\
&=&-\frac{N \Omega^2}{v^{D-2}} \delta_{l,m} \; \left( \frac{1}{D}-1 \right) \; \frac{2^{1-D} \pi^{1+ D/2}}{\Gamma\left( \frac{D}{2}\right) (2 \pi)^D} \; \sec\left( \frac{\pi D}{2}\right)
= -\frac{N \Omega^2}{12 \pi^2 v} \delta_{l,m} \left[ \frac{1}{\epsilon}\Omega^{-\epsilon}+ \frac{a}{2}  \right],
\end{eqnarray}
for $D=3-\epsilon$. Here, $D$ denotes the spatial dimensionality of the system and $N$ is the number of Weyl points. In the above expression $a=\left[5-3 \gamma_E + 3 \log(4\pi) \right]/3 \approx 3.62069$, where $\gamma_E \approx 0.577$ is the Euler-Mascheroni constant.

\subsection{Density-density correlator at $\Omega=0$}

The density-density correlation function for noninteracting Weyl fermions at a finite Matsubara frequency and a momentum  reads
\begin{equation}\label{eq:density correlator-1}
\Pi_{00}\left(i\Omega,{\bf q} \right) =- \int \frac{d^D {\mathbf k}}{(2\pi)^D} \; \int^{\infty}_{-\infty} \frac{d\omega}{2 \pi} \; { \rm \bf Tr} \;\left[\sigma_0\; G_f(i(\omega+\Omega),{\bf k}+{\bf q})\;\sigma_0\;  G_f(i\omega,{\bf k})\right].
\end{equation}

Explicit calculation for the  density-density correlator (with external frequency and momentum) yields
\begin{eqnarray}\label{eq:polarization-bubble-cond}
\Pi_{00}\left(i\Omega, {\mathbf q} \right) &=& - \int \frac{d^D {\mathbf k}}{(2\pi)^D} \; \int^{\infty}_{-\infty} \frac{d\omega}{2 \pi} \; {\mathbf {Tr}} \; \left[ \sigma_0 \frac{i (\omega+\Omega) + v {\boldsymbol \sigma} \cdot \left( {\mathbf k} +{\mathbf q} \right) }{(\omega+\Omega)^2 + v^2 \left( {\mathbf k} + {\mathbf q} \right)^2 } \; \sigma_0 \;  \frac{i \omega + v {\boldsymbol \sigma} \cdot {\mathbf k}}{\omega^2 + v^2 {\mathbf k}^2 } \right] \nonumber \\
&=& -2N \int \frac{d^D {\mathbf k}}{(2\pi)^D} \; \int^{\infty}_{-\infty} \frac{d\omega}{2 \pi} \: \frac{-\omega(\omega+\Omega)+ v^2 {\mathbf k} \cdot \left( {\mathbf k}+ {\mathbf q} \right)}{\left[ (\omega+\Omega)^2 + v^2 \left( {\mathbf k} + {\mathbf q} \right)^2 \right] \; \left[ \omega^2 + v^2 {\mathbf k}^2 \right]} \nonumber \\
&=& \frac{N}{v^D} \int \frac{d^D {\mathbf k}}{(2\pi)^D} \; \frac{|{\mathbf k}|+ |{\mathbf k}+{\mathbf q}|}{\Omega^2 + \left( |{\mathbf k}|+ |{\mathbf k}+{\mathbf q}| \right)^2} \: \left[1- \frac{{\mathbf k} \cdot \left( {\mathbf k} + {\mathbf q} \right) }{|{\mathbf k}|+ |{\mathbf k}+{\mathbf q}|} \right] =  \frac{N}{v^D} \int \frac{d^D {\mathbf k}}{(2\pi)^D} \frac{q^2 k^2 - \left( {\mathbf k} \cdot {\mathbf q} \right)^2}{k^3 \left( \Omega^2 + 4k^2 \right)} + {\mathcal O}\left( q^4 \right). \nonumber \\
\end{eqnarray}
While arriving at the final expression we perform Taylor series expansion in powers of $q$ and kept the terms only to the order $q^2$. In the intermediate step we have rescaled the momentum as $v {\mathbf k} \to {\mathbf k}$ and $v {\mathbf q} \to {\mathbf q}$. Now performing the integral over ${\mathbf k}$ (and restoring the factor of $v^2$ in front of $q^2$) we arrive at the final expression for the density-density correlator
\begin{eqnarray}\label{eq:polarization-bubble-cond-1}
\Pi_{00}\left(i\Omega, {\mathbf q} \right) = \frac{N}{v^D} \; \left( v^2 q^2 \right) \left( \frac{1}{D}-1 \right) \frac{2^{1-D} \pi^{1+ D/2}}{\Gamma\left( \frac{D}{2}\right) (2 \pi)^D} \; \sec\left( \frac{\pi D}{2}\right)= \frac{N q^2}{12 \pi^2 v} \left[ \frac{1}{\epsilon}q^{-\epsilon}+ \frac{a}{2}  \right],
\end{eqnarray}
where $D=3-\epsilon$.

We notice here that charge conservation
\begin{equation}
\partial_\tau \rho(\tau,{\bf r})+{\bm \nabla\cdot {\bf j}}(\tau,{\bf r})=0
 \end{equation}
 implies that  the polarization tensor satisfies
 \begin{equation}\label{CC-1}
 -i\Omega\Pi_{0\mu}(i\Omega,{\bf q})+q_l\Pi_{l\mu}(i\Omega,{\bf q})=0,
 \end{equation}
  and
 \begin{equation} \label{CC-2}
  \Omega^2\Pi_{00}(i\Omega,{\bf q})+q_l q_m \Pi_{lm}(i\Omega,{\bf q})=0,
 \end{equation}
   where index $\mu$ includes also the imaginary time, and summation over repeated indices is assumed.
Therefore, divergent parts of the current-current and the density-density correlation functions in Eqs. (\ref{eq:current-correlator-free}) and (\ref{eq:polarization-bubble-cond-1}) are consistent with charge conservation.
However, we will now compute the polarization function $\Pi(i\Omega, {\bf q})$ for a noninteracting Weyl at a finite frequency and momentum with $D=3-\epsilon$ regularization, and show  that the entire function at arbitrary momentum and frequency is consistent with charge conservation.

\subsection{Polarization tensor for noninteracting system}~\label{fullpolarization:nonint}

We start with the expression  for the components of the polarization tensor that includes Eqs.\ (\ref{eq:current correlator-1}) and (\ref{eq:density correlator-1}) as special cases
\begin{equation}\label{eq:full polarization}
\Pi_{\mu\nu}\left(i\Omega,{\bf q} \right) =- \int \frac{d^D {\mathbf k}}{(2\pi)^D} \; \int^{\infty}_{-\infty} \frac{d\omega}{2 \pi} \; { {\rm \bf Tr}} \;\left[\sigma_\mu\; G_f(i(\omega+\Omega),{\bf k}+{\bf q})\;\sigma_\nu\;  G_f(i\omega,{\bf k})\right]
\equiv -{\rm \bf Tr}(P_\mu(i\Omega,{\bf q})\sigma_\nu),
\end{equation}
where $\sigma_\mu=(\sigma_0,v {\bm \sigma})$.
Following the steps outlined in Appendix A of Ref.~\cite{herbut-juricic-vafek-2}, we then compute
\begin{equation}
P_\mu(i\Omega,{\bf q})=\int \frac{d^D {\mathbf k}}{(2\pi)^D} \; \int^{\infty}_{-\infty} \frac{d\omega}{2 \pi} \; { {\rm \bf Tr}}
\left[G_f(i\omega,{\bf k})\;\sigma_\mu\;G_f(i(\omega+\Omega),{\bf k}+{\bf q})\right].
\end{equation}
After introducing using the Feynman parameters, we obtain
\begin{align}
P_\mu(i\Omega,{\bf q})=\frac{1}{v^D}\int_0^1 dx\,\int\frac{d\omega}{2\pi}\,\int\frac{d^D{\bf k}}{(2\pi)^D}\,\frac{[i\omega+{\bm \sigma}\cdot{\bf k}]\sigma_\mu[i\omega+i\Omega+{\bm \sigma}\cdot({\bf k}+{\bf q})]}{[(\omega+x\Omega)^2+({\bf k}+x{\bf q})^2+\Delta]^2},
\end{align}
with $v{\bf q}\rightarrow {\bf q}$ and ${\Delta}=x(1-x)(\Omega^2+q^2)$. After integrating over $\omega$, shifting the momentum ${\bf k}+x{\bf q}\rightarrow {\bf k}$, and using the subsequent rotational symmetry of the integrand,  we arrive at the following expression
\begin{equation}
P_\mu(i\Omega,{\bf q})=\frac{1}{4v^D}\int_0^1dx\,\int\frac{d^D{\bf k}}{(2\pi)^D}\left[-\frac{\sigma_\mu}{\sqrt{k^2+\Delta}}+
\frac{(\sigma_l \sigma_\mu \sigma_l)k^2}{D(k^2+\Delta^2)^{3/2}}-\frac{x(1-x)(i\Omega+{\bm \sigma}\cdot{\bf q})\sigma_\mu (i\Omega+{\bm \sigma}\cdot{\bf q})}{(k^2+\Delta^2)^{3/2}}\right].
\end{equation}
After integrating over the momentum, we obtain
\begin{equation}
P_\mu(i\Omega,{\bf q})=\frac{F(D)}{v^D (\Omega^2+q^2)^{\frac{3-D}{2}}}\left[(\sigma_l\sigma_\mu\sigma_l-\sigma_\mu)(\Omega^2+q^2)+(D-1)(i\Omega+{\bm \sigma}\cdot{\bf q})\sigma_\mu (i\Omega+{\bm \sigma}\cdot{\bf q})\right],
\end{equation}
where
\begin{equation}
F(D)=\frac{\Gamma[\frac{1-D}{2}]\,\Gamma^2[\frac{D+1}{2}]}{4(4\pi)^{D/2}\,\Gamma[1/2]\,\Gamma[D+1]}.
\end{equation}
From here, using that $\sigma_l \sigma_l=D$, we find
\begin{equation}
P_0(i\Omega,{\bf q})=\frac{2F(D)(D-1)}{(\Omega^2+q^2)^{\frac{3-D}{2}}}\,[q^2+i\Omega {\bm\sigma}\cdot{\bf q}],
\end{equation}
which then yields for the components of the polarization
\begin{equation}
\Pi_{00}(i\Omega,{\bf q})=-\frac{4(D-1)F(D)}{v^{D-2}(\Omega^2+q^2)^{\frac{3-D}{2}}}q^2,
\end{equation}
and
\begin{equation}
\Pi_{0m}(i\Omega,{\bf q})=-\frac{4(D-1)F(D)}{v^{D-2}(\Omega^2+q^2)^{\frac{3-D}{2}}}i\Omega q_m,
\end{equation}
where we restored ${\bf q}\rightarrow v{\bf q}$.
To obtain $P_m(i\Omega,{\bf q})$, we have to recall that since we consider $D=3-\epsilon$, all three Pauli matrices should be used in the scalar products such as ${\bm\sigma}\cdot{\bf q}$ (unlike in $D=2$) which then yields
\begin{equation}
(i\Omega+{\bm\sigma}\cdot{\bf q})\sigma_m (i\Omega+{\bm\sigma}\cdot{\bf q})=(-\Omega^2+q^2)\sigma_m+2i\Omega q_m.
\end{equation}
Ultimately, we obtain
\begin{equation}
P_m(i\Omega,{\bf q})=\frac{2(D-1)F(D)}{v^{D-1}(\Omega^2+q^2)^{\frac{3-D}{2}}}(i\Omega q_m-\sigma_m \Omega^2),
\end{equation}
which then yields
\begin{equation}
\Pi_{lm}(i\Omega,{\bf q})=\frac{4(D-1)F(D)}{v^{D-2}(\Omega^2+q^2)^{\frac{3-D}{2}}}\Omega^2\delta_{lm},
\end{equation}
and
\begin{equation}
\Pi_{m0}(i\Omega,{\bf q})=-\frac{4(D-1)F(D)}{v^{D-2}(\Omega^2+q^2)^{\frac{3-D}{2}}}i\Omega q_m.
\end{equation}

It is now easy to check that the obtained polarization tensor is consistent with charge conservation, i.e. that it obeys Eqs. (\ref{CC-1}) and (\ref{CC-2}), and is therefore manifestly gauge invariant. Furthermore, taking $D=3-\epsilon$, we obtain the results for the ${\bf q}=0$ current-current and the $\Omega=0$ density-density correlators in Eqs.\ (\ref{eq:current-correlator-free}) and (\ref{eq:polarization-bubble-cond-1}).

\pagebreak

\subsection{Analytic continuation and optical conductivity}~\label{OC_Strategy}

We now proceed with the computation of the optical conductivity from the obtained polarization tensor. In order to extract the OC from the polarization tensor, we first identify the ultraviolet divergence appearing as $1/\epsilon$ as
\begin{equation}~\label{epsilontolambda}
\frac{1}{\epsilon} \equiv -\frac{1}{2} \; \log \left[1+\frac{4 v^2 \Lambda^2}{\Omega^2} \right],
\end{equation}
where $\Lambda$ is the ultraviolet momentum cutoff up to which the energy dispersion of Weyl fermions scales linearly with momentum. Now we perform the analytic continuation to real frequency according to $i \Omega \to \Omega +i \delta$ and subscribe to the definition of the real part of the optical conductivity
\begin{equation}~\label{OC:definition}
\sigma_{jj} = \left[ \frac{e^2_0}{h} \times 2 \pi \right] \: \lim_{\delta \to 0} \frac{\Im \left[ \Pi_{lm} \left( i \Omega \to \Omega +i \delta \right) \right]}{\Omega} = \frac{N e^2_0}{12 h v} \; \Omega \equiv \sigma_0 (\Omega).
\end{equation}
In the final expression we denote the OC in a noninteracting Weyl liquid as $\sigma_0$, and $e_0$ is the external test charge.


\section{Optical conductivity due to local density-density interaction}~\label{OC:short-range}

Now we systematically incorporate the correction to OC in a Weyl semimetal due to electron-electron interactions. For the sake of simplicity and to establish the methodology, we first focus on the short-range component of the density-density Coulomb interaction. It should be noted that in any lattice system the long range tail of the Coulomb interaction is always accompanied by its short-range component. Therefore, in  material-based and numerical experiments one needs to account for both components of the density-density interaction.

\subsection{Interaction correction to current-current correlator}

The correction to the current-current correlator due to the short-range component of density-density interaction (characterized by strength $g_0$) is given by $\delta \Pi_{lm} (i \Omega,0)=\delta \Pi^{SE}_{lm}(i \Omega,0) + \delta \Pi^{V}_{lm}(i \Omega,0)$. Respectively, $\delta \Pi^{SE}_{lm}(i \Omega,0)$ and $\delta \Pi^{V}_{lm}(i \Omega,0)$ accounts for self-energy and vertex diagrams. The contribution from the self-energy diagram reads as
\begin{eqnarray}
\delta\Pi^{SE}_{lm} (i \Omega,0) &=& (-1)^2 2 g_0 v^2 \int \frac{d^D {\mathbf k}}{(2\pi)^D} \; \int \frac{d^D {\mathbf p}}{(2\pi)^D} \; \int^{\infty}_{-\infty} \frac{d\omega}{2 \pi} \; \int^{\infty}_{-\infty} \frac{d\omega^\prime}{2 \pi} \; {\mathbf {Tr}} \bigg[ \frac{i \omega + v {\boldsymbol \sigma} \cdot {\mathbf k}}{\omega^2+v^2 {\mathbf k}^2} \; \sigma_l \; \frac{i (\omega+\Omega) + v {\boldsymbol \sigma} \cdot {\mathbf k}}{(\omega+\Omega)^2+v^2 {\mathbf k}^2} \; \frac{i \omega^\prime + v {\boldsymbol \sigma} \cdot {\mathbf p}}{(\omega^\prime)^2+v^2 {\mathbf p}^2} \; \nonumber \\
&\times& \; \frac{i (\omega+\Omega) + v {\boldsymbol \sigma} \cdot {\mathbf k}}{(\omega+\Omega)^2+v^2 {\mathbf k}^2}  \sigma_m \bigg] \; \equiv \; 0.
\end{eqnarray}
On the other hand, the contribution from the vertex diagram goes as
\begin{eqnarray}
\delta\Pi^{V}_{lm} (i \Omega,0) &=& (-1)^2 g_0 v^2 \int \frac{d^D {\mathbf k}}{(2\pi)^D} \; \int \frac{d^D {\mathbf p}}{(2\pi)^D} \; \int^{\infty}_{-\infty} \frac{d\omega}{2 \pi} \; \int^{\infty}_{-\infty} \frac{d\omega^\prime}{2 \pi} \; {\mathbf {Tr}} \bigg[  \frac{i \omega + v {\boldsymbol \sigma} \cdot {\mathbf k}}{\omega^2+v^2 {\mathbf k}^2} \; \sigma_l \; \frac{i (\omega+\Omega) + v {\boldsymbol \sigma} \cdot {\mathbf k}}{(\omega+\Omega)^2+v^2 {\mathbf k}^2} \; \frac{i \omega^\prime + v {\boldsymbol \sigma} \cdot {\mathbf p}}{(\omega^\prime)^2+v^2 {\mathbf p}^2}
\nonumber \\
&\times& \; \sigma_m \; \frac{i (\omega^\prime -\Omega) + v {\boldsymbol \sigma} \cdot {\mathbf p}}{(\omega^\prime-\Omega)^2+v^2 {\mathbf p}^2} \bigg] = g_0 v^2 \; {\mathbf {Tr}} \left[ I_{l} \left( i\Omega \right) \;  I_{m} \left( -i\Omega \right) \right],
\end{eqnarray}
where
\allowdisplaybreaks[4]
\begin{eqnarray}
I_l\left( i\Omega \right) &=& \int \frac{d^D {\mathbf k}}{(2\pi)^D} \; \int^{\infty}_{-\infty} \frac{d\omega}{2 \pi} \; \frac{i \omega + v {\boldsymbol \sigma} \cdot {\mathbf k}}{\omega^2+v^2 {\mathbf k}^2} \; \sigma_l \; \frac{i (\omega+\Omega) + v {\boldsymbol \sigma} \cdot {\mathbf k}}{(\omega+\Omega)^2+v^2 {\mathbf k}^2}
= \frac{\sigma_l}{v^D} \; \frac{2^2 \pi^{D/2} (1-D)}{D \Gamma\left( \frac{D}{2}\right)(2\pi)^D} \; \int^{\infty}_0 dk \frac{k^D}{4 k^2 + \Omega^2} \nonumber \\
&=& \sigma_l \frac{\Omega^2}{v^3} \: \left[\left(\frac{1}{D}-1 \right) \; \frac{2^{-1-D} \; \pi^{1+D/2}}{\Gamma\left( \frac{D}{2}\right) (2 \pi)^D}  \sec\left( \frac{\pi D}{2} \right)\right].
\end{eqnarray}
The factor of $(-1)^2$ in the correction arises because of the fermion loop, which gives a factor $-1$, and the Taylor expansion to the first order in the coupling, which also gives such a factor.
Therefore, the net contribution from the vertex diagram for $D=3-\epsilon$ reads as
\begin{eqnarray}\label{eq:current-short-range}
\delta\Pi^{V}_{lm} (i \Omega,0) = \delta_{l,m} \; 2N \; \frac{\Omega^4}{v^4} \; \left[\left(\frac{1}{D}-1 \right) \; \frac{2^{-D} \; \pi^{1+D/2}}{\Gamma\left( \frac{D}{2}\right) (2 \pi)^D}  \sec\left( \frac{\pi D}{2} \right)\right]^2 = \delta_{l,m} \; \left[ \frac{2N g_0 \Omega^4}{576 \pi^4 v^4} \right] \left[ \frac{1}{\epsilon^2} + \frac{a}{\epsilon} \right].
\end{eqnarray}

\subsection{Interaction correction to density-density correlator}

We now present the computation of the correction to the polarization tensor due to short-range interaction from density-density correlator. The contribution arising from the self-energy diagram is
\begin{eqnarray}
\delta \Pi^{SE}_{00} (i \Omega, {\mathbf q}) &=& (-1)^2 2 g_0 \int \frac{d^D {\mathbf k}}{(2\pi)^D} \; \int \frac{d^D {\mathbf p}}{(2\pi)^D} \; \int^{\infty}_{-\infty} \frac{d\omega}{2 \pi} \; \int^{\infty}_{-\infty} \frac{d\omega^\prime}{2 \pi} \; {\mathbf {Tr}} \bigg[ \frac{i \omega + v {\boldsymbol \sigma} \cdot {\mathbf k}}{\omega^2+v^2 {\mathbf k}^2} \; \sigma_0 \; \frac{i (\omega+\Omega) + v {\boldsymbol \sigma} \cdot \left( {\mathbf k} +{\mathbf q} \right)}{(\omega+\Omega)^2+v^2 \left( {\mathbf k}+ {\mathbf q} \right)^2}  \; \nonumber \\
&\times& \; \frac{i \omega^\prime + v {\boldsymbol \sigma} \cdot {\mathbf p}}{(\omega^\prime)^2+v^2 {\mathbf p}^2} \; \frac{i (\omega+\Omega) + v {\boldsymbol \sigma} \cdot \left( {\mathbf p} + {\mathbf q} \right)}{(\omega+\Omega)^2+v^2 \left( {\mathbf p} +{\mathbf q} \right)^2}  \sigma_0 \bigg] \; \equiv \; 0.
\end{eqnarray}
On the other hand, contribution from the vertex diagram reads as
\begin{eqnarray}
\delta \Pi^{V}_{00} (i \Omega, {\mathbf q}) &=& (-1)^2 g_0 {\mathbf {Tr}} \bigg\{ \int \frac{d^D {\mathbf k}}{(2\pi)^D} \;  \; \int^{\infty}_{-\infty} \frac{d\omega}{2 \pi} \; \bigg[  \frac{i \left( \omega -\Omega \right) + v {\boldsymbol \sigma} \cdot ({\mathbf k}-{\mathbf q})}{(\omega-\Omega)^2+v^2 ({\mathbf k}-{\mathbf q})^2} \; \sigma_0 \; \frac{i \omega + v {\boldsymbol \sigma} \cdot {\mathbf k}}{\omega^2+v^2 {\mathbf k}^2} \bigg] \nonumber \\
& \times & \int \frac{d^D {\mathbf p}}{(2\pi)^D} \; \int^{\infty}_{-\infty} \frac{d\omega^\prime}{2 \pi} \; \bigg[ \frac{i \omega^\prime + v {\boldsymbol \sigma} \cdot {\mathbf p}}{(\omega^\prime)^2+v^2 {\mathbf p}^2} \; \sigma_0 \; \frac{i \left( \omega^\prime -\Omega \right) + v {\boldsymbol \sigma} \cdot ({\mathbf p}-{\mathbf q})}{(\omega^\prime-\Omega)^2+v^2 ({\mathbf p}-{\mathbf q})^2}   \bigg] \bigg\}.
\end{eqnarray}
Note that for short-range interaction the contribution from the vertex diagram breaks into two pieces. We now show some essential steps of the analysis, which will also be useful while we compute the same diagram, but in the presence of the long-range tail of the Coulomb interaction. At this stage we first rescale the momentum according to $v {\mathbf k} \to {\mathbf k}$, $v {\mathbf p} \to {\mathbf p}$, $v {\mathbf q} \to {\mathbf q}$. Note that
\begin{eqnarray}~\label{I1_Vertex}
&&I_{1}({\mathbf p}, {\mathbf q}, \Omega)=\int^{\infty}_{-\infty} \; \frac{d \omega^\prime}{2 \pi}  \: \frac{i \omega^\prime +  {\boldsymbol \sigma} \cdot {\mathbf p}}{(\omega^\prime)^2+ {\mathbf p}^2} \: \frac{i \left( \omega^\prime -\Omega \right) + {\boldsymbol \sigma} \cdot ({\mathbf p}-{\mathbf q})}{(\omega^\prime-\Omega)^2+ ({\mathbf k}-{\mathbf q})^2} = \frac{1}{2 \left[ \Omega^2 + \left(p+|{\mathbf p}-{\mathbf q}| \right)^2 \right]} \nonumber \\
&\times&  \; \bigg[
- \left( p+|{\mathbf p}-{\mathbf q}| \right) \left[ 1 - \frac{{\mathbf p} \cdot ({\mathbf p}-{\mathbf q})}{p |{\mathbf p}-{\mathbf q}|}\right]
+ i {\boldsymbol \sigma} \cdot \left\{ \Omega  \left( \frac{{\mathbf p}-{\mathbf q}}{|{\mathbf p}-{\mathbf q}|} - \frac{{\mathbf p}}{p} \right)
-   \left( {\mathbf p} \times {\mathbf q} \right) \; \frac{p+|{\mathbf p}-{\mathbf q}| }{p |{\mathbf p}-{\mathbf q}|}  \right\} \bigg]
\equiv a + {\boldsymbol \sigma} \cdot {\mathbf b}
\end{eqnarray}
Similarly, after completing the integral over Matsubara frequency $\omega$ we can write
\begin{equation}~\label{I2_Vertex}
I_{2}({\mathbf p}, {\mathbf q}, \Omega)= \int^{\infty}_{-\infty} \frac{d\omega}{2 \pi} \;  \frac{i \left( \omega -\Omega \right) + {\boldsymbol \sigma} \cdot ({\mathbf k}-{\mathbf q})}{(\omega-\Omega)^2+ ({\mathbf k}-{\mathbf q})^2} \;  \frac{i \omega + {\boldsymbol \sigma} \cdot {\mathbf k}}{\omega^2+ {\mathbf k}^2} \equiv c + {\boldsymbol \sigma} \cdot {\mathbf d},
\end{equation}
with
\begin{eqnarray}~\label{abcd}
a &=&  -\frac{p+|{\mathbf p}-{\mathbf q}|}{2 \left[ \Omega^2 + \left(p+|{\mathbf p}-{\mathbf q}| \right)^2 \right]} \left[ 1 - \frac{{\mathbf p} \cdot ({\mathbf p}-{\mathbf q})}{p |{\mathbf p}-{\mathbf q}|}\right], \:
b=  \frac{i \Omega  \left( \frac{{\mathbf p}-{\mathbf q}}{|{\mathbf p}-{\mathbf q}|} - \frac{{\mathbf p}}{p} \right)
-   i \left( {\mathbf p} \times {\mathbf q} \right) \; \left( \frac{1}{|{\mathbf p}-{\mathbf q}|} + \frac{1}{p} \right)}  {2 \left[ \Omega^2 + \left(p+|{\mathbf p}-{\mathbf q}| \right)^2 \right]}, \nonumber \\
c &=&   -\frac{k+|{\mathbf k}+{\mathbf q}|}{2 \left[ \Omega^2 + \left(k+|{\mathbf k}+{\mathbf q}| \right)^2 \right]} \left[ 1 - \frac{{\mathbf k} \cdot ({\mathbf k}+{\mathbf q})}{k |{\mathbf k}+{\mathbf q}|}\right], \:
d= \frac{i \Omega  \left( \frac{{\mathbf k}+{\mathbf q}}{|{\mathbf k}+{\mathbf q}|} - \frac{{\mathbf k}}{k} \right)
+   i \left( {\mathbf k} \times {\mathbf q} \right) \; \left( \frac{1}{|{\mathbf k}+{\mathbf q}|} + \frac{1}{k} \right) }{2 \left[ \Omega^2 + \left(k+|{\mathbf k}+{\mathbf q}| \right)^2 \right]}.
\end{eqnarray}
In terms of above parameters $\delta \Pi^{V}_{00} (i \Omega, {\mathbf q})$ can be written as
\begin{eqnarray}
\delta \Pi^{V}_{00} (i \Omega, {\mathbf q}) =  \frac{g_0}{v^6}  \int \frac{d^D {\mathbf k}}{(2\pi)^D} \int \frac{d^D {\mathbf p}}{(2\pi)^D} \;
{\mathbf {Tr}} \left[ \left( a + {\boldsymbol \sigma} \cdot {\mathbf b} \right) \left( c + {\boldsymbol \sigma} \cdot {\mathbf d} \right) \right]=- \frac{2N g_0}{v^6} \int \frac{d^D {\mathbf k}}{(2\pi)^D} \int \frac{d^D {\mathbf p}}{(2\pi)^D} \left[ ac + {\mathbf b} \cdot {\mathbf d} \right].
\end{eqnarray}
Since we are interested in contribution proportional to $q^2$, next we expand all these quantities to the order $q^2$, yielding
\begin{eqnarray}~\label{abcd_expansion}
a &=& \frac{4 ({\mathbf p} \cdot {\mathbf q})^2+p^2 q^2}{p^3 \left(\Omega^2 +4 p^2 \right)}, \quad
b=i \; \frac{ \Omega[ {\mathbf p} ({\mathbf p} \cdot {\mathbf q})- p^2 {\mathbf q} ]-2 p^2 ({\mathbf p} \times {\mathbf q})}{p^3 \left(\Omega^2 +4 p^2 \right)}, \nonumber \\
c &=& \frac{4 ({\mathbf k} \cdot {\mathbf q})^2+k^2 q^2}{k^3 \left(\Omega^2 +4 k^2 \right)}, \quad
d=i \; \frac{ \Omega[ {\mathbf k} ({\mathbf k} \cdot {\mathbf q})- k^2 {\mathbf q} ]+2 k^2 ({\mathbf k} \times {\mathbf q})}{k^3 \left(\Omega^2 +4 k^2 \right)}.
\end{eqnarray}
Notice that the product $ac \sim {\mathcal O}(q^4)$ and therefore does not contribute to the conductivity. We can compactly write
\begin{eqnarray}
{\mathbf b} \cdot {\mathbf d}=-\Omega^2 \frac{ \left[ {\mathbf p} \left( {\mathbf p} \cdot {\mathbf q} \right) -p^2 {\mathbf q} \right] \cdot \left[ {\mathbf k} \left( {\mathbf k} \cdot {\mathbf q} \right) -k^2 {\mathbf q} \right]  }{ 4 k^3 p^3 \left(\Omega^2 +4 k^2 \right) \left(\Omega^2 +4 p^2 \right)}
- \frac{ \left( {\mathbf p} \times {\mathbf q} \right) \cdot \left( {\mathbf k} \times {\mathbf q} \right) }{k p \left(\Omega^2 +4 k^2 \right) \left(\Omega^2 +4 p^2 \right)}
- \frac{\Omega {\mathbf q} \cdot ({\mathbf k} \times {\mathbf p})}{2 k^3 p^3} \frac{\left[ p^2 ({\mathbf k} \cdot {\mathbf q} ) + k^2 ({\mathbf p} \cdot {\mathbf q} ) \right]}{\left(\Omega^2 +4 k^2 \right) \left(\Omega^2 +4 p^2 \right)}. \nonumber \\
\end{eqnarray}
As the last term is \emph{odd} under the exchange of ${\mathbf k}$ and ${\mathbf p}$, it does not contribute after the momentum integral. Also the two pieces in the second term in the last expression are individually odd functions of ${\mathbf k}$ and ${\mathbf p}$, and therefore both vanish. After these simplifications the net contribution from the vertex diagram goes as
\begin{eqnarray}\label{eq:density-short-range}
\delta \Pi^{V}_{00} (i \Omega, {\mathbf q}) &=& -\frac{N g_0 \Omega^2}{2 v^6} \: \left[ \int \frac{d^D {\mathbf k}}{(2\pi)^D} \:
\frac{{\mathbf k} ({\mathbf k} \cdot {\mathbf q}) -k^2 {\mathbf q}}{k^3 \left(\Omega^2 +4 k^2 \right)} \right] \cdot
\left[ \int \frac{d^D {\mathbf p}}{(2\pi)^D} \:
\frac{{\mathbf p} ({\mathbf p} \cdot {\mathbf q}) -p^2 {\mathbf q}}{p^3 \left(\Omega^2 +4 k^2 \right)} \right] \nonumber \\
&=& -\frac{2 N g_0 \Omega^2 \left(v^2 q^2\right)}{ v^6} \left[\left(\frac{1}{D}-1 \right) \; \frac{2^{-D} \; \pi^{1+D/2}}{\Gamma\left( \frac{D}{2}\right) (2 \pi)^D}  \sec\left( \frac{\pi D}{2} \right)\right]^2
= -\left[ \frac{2N g_0 \Omega^2 q^2}{576 \pi^4 v^4} \right] \left[ \frac{1}{\epsilon^2} + \frac{a}{\epsilon} \right].
\end{eqnarray}

Notice that the results for the interaction correction to the current-current and the density-density correlators in Eqs. (\ref{eq:current-short-range}) and (\ref{eq:density-short-range}), respectively, are consistent with the
charge conservation, Eq. (\ref{CC-2}).

\subsection{Optical conductivity}

From expression for the polarization tensor, now identifying the ultraviolet divergence following Eq.~(\ref{epsilontolambda}) and following the definition of the real part of optical conductivity [see Eq.~(\ref{OC:definition})], we find the correction of the OC due to the short-range component of the density-density interaction to be
\begin{equation}
\delta\sigma_{jj}(\Omega)=\sigma_0 (\Omega) \; \left[ \frac{g_0 \Omega^2}{24 \pi^2 v^3}\right] \;\left\{ a- 2 \log\left( \frac{E_\Lambda}{ \Omega}\right) \right\},
\end{equation}
where $E_\Lambda=2 \Lambda v$ and the quantity inside the straight bracket ``$[]$" is the \emph{dimensionless short-range coupling constant} in $D=3$.
\\


\section{Optical conductivity due to long-range tail of the Coulomb interaction}~\label{OC:long-range}

Finally, we turn our focus to the computation of the correction to the OC due to the long-range tail of the Coulomb interaction. This is the most challenging part of the analysis. Nonetheless, we have already developed a vast amount of technical aspect of this problem to facilitate the discussion in this section. Once again we will compute the polarization tensor from (a) current-current correlator and then (b) density-density correlator. And finally from either of these two expressions we will compute the correction to the OC.

\subsection{Current-current correlator}

The correction to the current correlator has the self-energy and the vertex parts, which we compute separately.

Contribution to the current-current correlator, for which we consider only one of the diagonal components due to isotropy, at zero momentum arising from the self-energy diagram is of the form
\begin{eqnarray}
\delta \Pi^{SE}_{xx}(i\Omega,0) &=&-i^2 (2N)\int\frac{d\omega}{2\pi}\int\frac{d\omega'}{2\pi}\int \frac{d^D {\mathbf k}}{(2\pi)^D}\int \frac{d^D {\mathbf p}}{(2\pi)^D}{\rm \bf Tr}[G_f(i\omega,{\bf k})\, v\sigma_x\, G_f(i\omega+i\Omega,{\bf k}) \nonumber \\
&\times& \,v\sigma_x\,G_f(i\omega,{\bf k})\,G_f(i\omega',{\bf p})\, V_C({\bf k}-{\bf p})].
\end{eqnarray}
Here, a fermion loop gives a factor of $-1$, while the Coulomb vertex which appears twice to the leading order produces the factor $i^2$. The  integrals over the Matsubara frequencies are performed from $-\infty$ to $\infty$, i.e. $\int d\omega\equiv \int_{-\infty}^\infty d\omega$. Explicit form of this contribution then reads
\begin{eqnarray}
\delta \Pi^{SE}_{xx}&=&-\frac{2N}{v^2} {\mathbf {Tr}} \int \frac{d^D {\mathbf k}}{(2\pi)^D}  \int \frac{d\omega}{2 \pi}  \frac{i \omega +  {\boldsymbol \sigma} \cdot {\mathbf k}}{\omega^2+ {\mathbf k}^2} \sigma_x \frac{i (\omega+\Omega) +  {\boldsymbol \sigma} \cdot {\mathbf k}}{(\omega+\Omega)^2+ {\mathbf k}^2} \sigma_x \frac{i \omega +  {\boldsymbol \sigma} \cdot {\mathbf k}}{\omega^2+ {\mathbf k}^2}
\left[ i^2\int \frac{d^D {\mathbf p}}{(2\pi)^D} \int \frac{d\omega^\prime}{2 \pi} \frac{i \omega^\prime +  {\boldsymbol \sigma} \cdot {\mathbf p}}{\left(\omega^\prime\right)^2+ {\mathbf p}^2} \frac{2 \pi e^2}{|{\mathbf k}-{\mathbf p}|^2} \right] \nonumber \\
&=& -\frac{2N}{v^2} {\mathbf {Tr}} \left\{ \int \frac{d^D {\mathbf k}}{(2\pi)^D}  \int^{\infty}_{-\infty} \frac{d\omega}{2 \pi}  \frac{i \omega +  {\boldsymbol \sigma} \cdot {\mathbf k}}{\omega^2+ {\mathbf k}^2} \sigma_x \frac{i (\omega+\Omega) +  {\boldsymbol \sigma} \cdot {\mathbf k}}{(\omega+\Omega)^2+ {\mathbf k}^2} \sigma_x \frac{i \omega +  {\boldsymbol \sigma} \cdot {\mathbf k}}{\omega^2+ {\mathbf k}^2} \Sigma\left( {\mathbf k}\right) \right\},
\end{eqnarray}
where $ \Sigma({\mathbf k})$ is the self-energy correction due to long-range interaction, given by Eq.\ (\ref{eq:self-energy}), which for convenience we write again
\begin{eqnarray}\label{eq:self-energy-1}
 \Sigma({\mathbf k})=-\frac{\pi e^2}{2} \: \left[ \frac{\Gamma\left( \frac{3-D}{2}\right) \Gamma\left( \frac{D}{2}-1\right) \Gamma\left( \frac{D+1}{2}\right)}{(4 \pi)^{D/2} \Gamma\left(\frac{3}{2} \right) \Gamma \left(D-\frac{1}{2} \right)} \right] \: \frac{{\boldsymbol \sigma} \cdot {\mathbf k}}{|{\mathbf k}|^{3-D}} \equiv -E(D) \: \frac{{\boldsymbol \sigma} \cdot {\mathbf k}}{|{\mathbf k}|^{3-D}}.
\end{eqnarray}
After performing the trace algebra we arrive at the following compact expression for the self-energy diagram
\begin{eqnarray}
\delta \Pi^{SE}_{xx}=4 N E(D) \int \frac{d^D {\mathbf k}}{(2\pi)^D}  \int \frac{d\omega}{2 \pi} \: \frac{ \left[ (\omega^2-k^2)(k^2-2k^2_x) -2\omega(\omega+\Omega)k^2 \right]}{k^{3-D} \left[ \omega^2+k^2 \right]^2 \left[(\omega+\Omega)^2+ k^2 \right]}.
\end{eqnarray}
Performing the integral over the Matsubara frequency $\omega$ we obtain
\begin{eqnarray}\label{eq:current-selfenergy-final}
\delta \Pi^{SE}_{xx} &=& -\frac{2N \pi e^2}{v^2} \left[ \frac{\Gamma\left( \frac{3-D}{2}\right) \Gamma\left( \frac{D}{2}-1\right) \Gamma\left( \frac{D+1}{2}\right)}{(4 \pi)^{D/2} \Gamma\left(\frac{3}{2} \right) \Gamma \left(D-\frac{1}{2} \right)} \right] \left( 1-\frac{1}{D}\right) \: \int \frac{d^D {\mathbf k}}{(2\pi)^D} \frac{k^{D-2} \left(4k^2-\Omega^2 \right)}{\left(\Omega^2+4k^2 \right)^2} \nonumber \\
&=&  \frac{2 N \pi e^2 \Omega^2}{v^2} \left[ \frac{\Gamma\left( \frac{3-D}{2}\right) \Gamma\left( \frac{D}{2}-1\right) \Gamma\left( \frac{D+1}{2}\right)}{(4 \pi)^{D/2} \Gamma\left(\frac{3}{2} \right) \Gamma \left(D-\frac{1}{2} \right)} \frac{2 \pi^{D/2}}{4 \Gamma\left( \frac{D}{2}\right) (2\pi)^D} \left( 1-\frac{1}{D}\right) \right] \int^{\infty}_0 dk \; \frac{k^{2D-5} \left( \Omega^2+12 k^2\right)}{\left(\Omega^2+4k^2 \right)^2} \nonumber \\
&=& \frac{2 N \pi e^2 \Omega^2}{v^2} \left[ \frac{\Gamma\left( \frac{3-D}{2}\right) \Gamma\left( \frac{D}{2}-1\right) \Gamma\left( \frac{D+1}{2}\right)}{(4 \pi)^{D/2} \Gamma\left(\frac{3}{2} \right) \Gamma \left(D-\frac{1}{2} \right)} \frac{2 \pi^{D/2}}{4 \Gamma\left( \frac{D}{2}\right) (2\pi)^D} \left( 1-\frac{1}{D}\right) \right] 2^{3-2D} \pi (2D-3) \; \mathrm{cosec}(\pi D) \nonumber \\
&=& \frac{N e^2 \Omega^2}{72 \pi^3 v^2} \; \left[ \frac{3}{\epsilon^2} + \frac{1}{\epsilon} \left[ 7-3 \gamma_E + 3 \log(4 \pi)\right]\right],
\end{eqnarray}
for $D=3-\epsilon$. We point out that while arriving to the second line, we subtract the $\Omega=0$ piece of the bubble.
\\

Now we turn our attention to the vertex diagram. The expression for the polarization bubble arising from this diagram reads as
\begin{eqnarray}
\delta \Pi^{V}_{xx} (i \Omega,0) &=& -i^2 N \int \frac{d^D {\mathbf k}}{(2\pi)^D} \int \frac{d^D {\mathbf p}}{(2\pi)^D} \int \frac{d\omega}{2 \pi}
\int \frac{d\omega^\prime}{2 \pi}{\rm \bf Tr}[G_f(i\omega,{\bf k})\,v\sigma_x\,G_f(i\omega+i\Omega,{\bf k})\,G_f(i\omega+i\Omega,{\bf p}) \nonumber \\
&\times& v\sigma_x\, G_f(i\omega,{\bf p})V_C({\bf k}-{\bf p})].
\end{eqnarray}
Again, a fermion loop gives a factor of $-1$, while the Coulomb vertex which appears twice to the leading order produces the factor $i^2$.
We now explicitly write this expression as
\begin{eqnarray}
\delta \Pi^{V}_{xx} (i \Omega,0)&=&-i^2\frac{N}{v^2} \int \frac{d^D {\mathbf k}}{(2\pi)^D} \int \frac{d^D {\mathbf p}}{(2\pi)^D} \int \frac{d\omega}{2 \pi} \int\frac{d\omega^\prime}{2 \pi} {\rm \bf Tr} \bigg\{ \frac{i \omega +  {\boldsymbol \sigma} \cdot {\mathbf k}}{\omega^2+ {\mathbf k}^2} \sigma_x \frac{i (\omega+\Omega) +  {\boldsymbol \sigma} \cdot {\mathbf k}}{(\omega+\Omega)^2+ {\mathbf k}^2}\; \frac{i (\omega+\Omega) +  {\boldsymbol \sigma} \cdot {\mathbf p}}{(\omega+\Omega)^2+ {\mathbf p}^2} \nonumber \\
&\times& \sigma_x \frac{i \omega^\prime +  {\boldsymbol \sigma} \cdot {\mathbf p}}{\left( \omega^\prime \right)^2+ {\mathbf k}^2}
\bigg\} \; \frac{2 \pi e^2}{|{\mathbf k}-{\mathbf p}|^2}.
\end{eqnarray}
After performing the trace algebra and completing the frequency integrals using the residue technique we find
\begin{eqnarray}
\delta \Pi^{V}_{xx} (i \Omega,0)&=&\frac{4 \pi N e^2}{v^2}  \int \frac{d^D {\mathbf k}}{(2\pi)^D} \int \frac{d^D {\mathbf p}}{(2\pi)^D} \;
\frac{\Omega^2 \left(p_x k_x -{\mathbf k} \cdot {\mathbf p} \right) + 4 \left( k_x p_x {\mathbf k} \cdot {\mathbf p} + k^2 p^2 -p^2_x k^2-k^2_x p^2 \right)}{k p \left( \Omega^2 +4k^2 \right) \left( \Omega^2 +4p^2 \right) |{\mathbf k}-{\mathbf p}|^2}.
\end{eqnarray}
Now we subtract the zero frequency piece of $\delta \Pi^{V}_{xx} (i \Omega,0)$, given by
\begin{eqnarray}
\delta \Pi^{V}_{xx} (0,0)= \frac{4 \pi N e^2}{v^2} \int \frac{d^D {\mathbf k}}{(2\pi)^D} \int \frac{d^D {\mathbf p}}{(2\pi)^D} \frac{4 \left( k_x p_x {\mathbf k} \cdot {\mathbf p} + k^2 p^2 -p^2_x k^2-k^2_x p^2  \right) }{16 k^3 p^3 |{\mathbf k}-{\mathbf p}|^2},
\end{eqnarray}
to arrive at the following compact expression for the vertex diagram
\allowdisplaybreaks[4]
\begin{eqnarray}
&&\delta \Pi^{V}_{xx} (i \Omega,0) \nonumber \\
&=&-\frac{4 \pi N e^2 \Omega^2}{v^2} \int \frac{d^D {\mathbf k}}{(2\pi)^D} \int \frac{d^D {\mathbf p}}{(2\pi)^D}
\frac{4 k^2 p^2 \left( {\mathbf k} \cdot {\mathbf p} -p_x k_x \right) + \left[ \Omega^2 + 4\left( k^2 +p^2 \right)\right] \left( k_x p_x {\mathbf k} \cdot {\mathbf p} + k^2 p^2 -p^2_x k^2-k^2_x p^2  \right)}{4 k^3 p^3 \left( \Omega^2+4p^2 \right) \left( \Omega^2+4k^2 \right) \; |{\mathbf k}-{\mathbf p}|^2} \nonumber \\
&=& -\frac{1}{v^2}\left[ \delta \Pi^{V,1}_{xx,1} (i \Omega,0) + \delta \Pi^{V,2}_{xx,1} (i \Omega,0) + \delta \Pi^{V,3}_{xx,1} (i \Omega,0) + \delta \Pi^{V}_{xx,2} (i \Omega,0) \right].
\end{eqnarray}
Various pieces appearing in the above expression read as
\allowdisplaybreaks[4]
\begin{eqnarray}
\delta \Pi^{V,1}_{xx,1} (i \Omega,0) &=& \frac{N\Omega^2}{4} \int \frac{d^D {\mathbf k}}{(2\pi)^D} \int \frac{d^D {\mathbf p}}{(2\pi)^D} \frac{2 \pi e^2}{|{\mathbf k}-{\mathbf p}|^2} \; \frac{k_x p_x {\mathbf k} \cdot {\mathbf p} + k^2 p^2 -p^2_x k^2-k^2_x p^2 }{k^3 p^3 \left[ k^2 + (\Omega/2)^2 \right]}, \label{eq:current-vertex-11}\\
\delta \Pi^{V,2}_{xx,1} (i \Omega,0) &=& -\frac{N\Omega^4}{32} \int \frac{d^D {\mathbf k}}{(2\pi)^D} \int \frac{d^D {\mathbf p}}{(2\pi)^D} \frac{2 \pi e^2}{|{\mathbf k}-{\mathbf p}|^2} \frac{1}{k p \left[ k^2 + (\Omega/2)^2 \right] \left[ p^2 + (\Omega/2)^2 \right] },\label{eq:current-vertex-21} \\
\delta \Pi^{V,3}_{xx,1} (i \Omega,0) &=& -\frac{N\Omega^4}{32} \int \frac{d^D {\mathbf k}}{(2\pi)^D} \int \frac{d^D {\mathbf p}}{(2\pi)^D} \frac{2 \pi e^2}{|{\mathbf k}-{\mathbf p}|^2}  \frac{k_x p_x {\mathbf k} \cdot {\mathbf p} - 2 p^2 k^2_x}{k^3 p^3 \left[ k^2 + (\Omega/2)^2 \right] \left[ p^2 + (\Omega/2)^2 \right]}, \label{eq:current-vertex-31}\\
\delta \Pi^{V}_{xx,2} (i \Omega,0) &=& \frac{N\Omega^2}{8} \int \frac{d^D {\mathbf k}}{(2\pi)^D} \int \frac{d^D {\mathbf p}}{(2\pi)^D} \frac{2 \pi e^2}{|{\mathbf k}-{\mathbf p}|^2} \frac{{\mathbf k} \cdot {\mathbf p} -p_x k_x}{k p \left[ k^2 + (\Omega/2)^2 \right] \left[ p^2 + (\Omega/2)^2 \right]}.\label{eq:current-vertex-2}
\end{eqnarray}
Now we present some details of the computation of each term.
\\

Note that following identity involving Feynman parameters is extremely useful to compute these terms
\begin{equation}
\frac{1}{A^\alpha \; B^\beta \; C^\gamma} = \frac{\Gamma\left(\alpha+\beta+\gamma \right)}{\Gamma \left( \alpha \right) \Gamma \left( \beta \right) \Gamma \left( \gamma \right)} \:\:  \int^1_0 dx \int^{1-x}_0 \: dy \:\: \frac{(1-x-y)^{\alpha-1} x^{\beta-1} y^{\gamma-1}}{\left[ (1-x-y)A + x B + y C \right]^{\alpha+\beta+\gamma}}.
\end{equation}
Now the term $\delta \Pi^{V,1}_{xx,1} (i \Omega,0)$ can be written as
\begin{eqnarray}\label{eq:current-vertex-11-final}
&&\delta \Pi^{V,1}_{xx,1} (i \Omega,0) = \frac{N\Omega^2}{4} (2\pi e^2) \frac{\Gamma\left( \frac{7}{2}\right)}{\Gamma\left( \frac{3}{2}\right)} \int^1_0 dx \int^{1-x}_0 dy \int \frac{d^D {\mathbf p}}{(2\pi)^D} \frac{p^2-p^2_x}{p^3} \int\frac{d^D {\mathbf k}}{(2\pi)^D} \frac{\left[ \left( 1-\frac{1}{D}\right) k^2 + x^2 p^2 \right] \left(1-x-y\right)^{1/2}}{\left[ k^2 + x(1-x) p^2 + y (\Omega/2)^2 \right]^{7/2}} \nonumber \\
&=& \frac{N\Omega^2}{4} (2\pi e^2) \frac{\left(1-\frac{1}{D}\right) \Gamma\left( \frac{5-D}{2}\right)}{\Gamma\left( \frac{3}{2}\right) (4\pi)^{D/2}} \frac{2 \pi^{D/2}}{\Gamma \left(\frac{D}{2} \right) (2 \pi)^{D}}\int^1_0 dx \int^{1-x}_0 dy \int^{\infty}_0 dp \frac{p^{D-2} \sqrt{1-x-y}}{\left[ x(1-x)p^2+y(\Omega/2)^2\right]^{\frac{5-D}{2}}} \bigg[ \frac{D-1}{2} \nonumber \\
&+& \frac{\frac{5-D}{2} x^2 p^2}{x(1-x)p^2+y(\Omega/2)^2} \bigg]
=N\Omega^2 e^2 \frac{\pi^{1+\frac{D}{2}} \Gamma(3-D) \Gamma\left( \frac{1+D}{2}\right) \left(1-\frac{1}{D} \right)}{\Gamma\left( \frac{3}{2}\right) \Gamma\left(\frac{D}{2} \right) (2 \pi)^{D} (4\pi)^{D/2} }\int^1_0 dx \frac{x^{\frac{1-D}{2}}}{\left(1-x\right)^{\frac{D+1}{2}}}  \int^{1-x}_0 dy y^{D-3} \sqrt{1-x-y} \nonumber \\
&=& N\Omega^2 e^2 \frac{\pi^{1+\frac{D}{2}} \Gamma(3-D) \Gamma\left( \frac{1+D}{2}\right) \left(1-\frac{1}{D} \right) \Gamma\left(\frac{3-D}{2}\right) \Gamma\left(\frac{D}{2}-1 \right)}{\Gamma\left( 3/2 \right) \Gamma\left(D/2 \right) (2 \pi)^{D} (4\pi)^{D/2} \Gamma\left( 1/2 \right) } \; \Gamma(D-2)  =\frac{Ne^2 \Omega^2}{72 \pi^3} \left[ \frac{2}{\epsilon^2} + \frac{6-2\gamma_E+2 \log(4\pi)}{\epsilon}\right].
\end{eqnarray}
Next we compute $\delta \Pi^{V,2}_{xx,1} (i \Omega,0)$ given by
\begin{eqnarray}
&&\delta \Pi^{V,2}_{xx,1} (i \Omega,0) \nonumber \\
&=&-2 \pi e^2 \frac{N\Omega^4}{32} \frac{\Gamma\left( \frac{5}{2}\right)}{\Gamma\left( \frac{1}{2}\right)} \int^1_0 dx \int^{1-x}_0 \frac{dy}{\sqrt{1-x-y}} \int \frac{d^D {\mathbf k}}{(2\pi)^D} \frac{1}{k \left[ k^2 + (\Omega/2)^2 \right]} \int \frac{d^D {\mathbf p}}{(2\pi)^D} \frac{1}{\left[ p^2+x(1-x)k^2 +y(\Omega/2)^2 \right]^{5/2}} \nonumber \\
&=&  -2 \pi e^2 \frac{N \Omega^4}{32} \frac{2 \pi^{D/2} \Gamma\left(\frac{5-D}{2} \right)}{\Gamma\left(\frac{D}{2} \right) (2 \pi)^D \Gamma(1/2) (4\pi)^{D/2}} \int^1_0 dx \int^{1-x}_0 dy \int^\infty_0 dk \; \frac{k^{D-2} \left[ 1-x-y \right]^{-1/2} }{\left[ k^2 + (\Omega/2)^2 \right] \left[ x(1-x) k^2 + y(\Omega/2)^2 \right]} \nonumber \\
&=& Ne^2\Omega^{2D-4}  \frac{2^{4-2D} \pi^{\frac{1+D}{2}}}{(2\pi)^D (4\pi)^{D/2} \Gamma(D/2)} \int^1_0 dx \int^{1-x}_0 \frac{dy}{\sqrt{1-x-y}} \bigg[ \frac{\pi}{\sin(\pi D)} \Gamma\left( \frac{5-D}{2}\right) \left[ x(1-x)\right]^{\frac{D-5}{2}} \nonumber \\
&-& \Gamma(3-D) \Gamma\left( \frac{D-1}{2}\right) y^{D-3} \left[ x(1-x)\right]^{\frac{1-D}{2}} {_2}F_1 \left[ 1,\frac{D-1}{2},D-2,\frac{y}{x(1-x)}\right] \bigg]= - N e^2 \Omega^2 \; \frac{7 \; \zeta \left( 3\right)}{128 \pi^2}.
\end{eqnarray}
Here, ${_2}F_1(a,b,c,z)$ is the ordinary hypergeometric function. Next we compute $\delta \Pi^{V,3}_{xx,1} (i \Omega,0)$ which can be expressed as
\allowdisplaybreaks[4]
\begin{eqnarray}
&& \delta \Pi^{V,3}_{xx,1} (i \Omega,0) \nonumber \\
&=& - \pi e^2 \frac{N \Omega^4}{16} \frac{\Gamma\left( \frac{7}{2}\right)}{\Gamma\left( \frac{3}{2}\right)} \int^1_0 dx \int^{1-x}_0 dy \int \frac{d^D {\mathbf p}}{(2\pi)^D} \frac{\left[ 1-x-y\right]^{-1/2}}{p^3 \left[ p^2 + \left(\frac{\Omega}{2} \right)^2 \right]} \int \frac{d^D {\mathbf k}}{(2\pi)^D} \frac{\frac{1}{D} \left[k^2p^2_x -2 p^2 k^2 \right] -x^2 p^2 p^2_x  }{\left[ k^2+ x(1-x)p^2+y \left(\frac{\Omega}{2} \right)^2 \right]^{7/2}} \nonumber \\
&=& - \pi e^2 \frac{N \Omega^4}{16} \frac{2 \pi^{D/2} \Gamma\left( \frac{5-D}{2} \right)}{(2\pi)^D (4\pi)^{D/2} \Gamma(D/2)} \int^1_0 dx \int^{1-x}_0 dy \int^\infty_0 dp \; \frac{p^{D-2} \left[ 1-x-y\right]^{-1/2} \left[ x(1-x) \right]^{\frac{D-5}{2}}}{\left[ p^2+ \left(\frac{\Omega}{2} \right)^2 \right] \left[ p^2+ \frac{y}{x(1-x)}\left(\frac{\Omega}{2} \right)^2 \right]^{\frac{5-D}{2}}} \nonumber \\
&\times& \left[ \frac{1}{2} \left( \frac{1}{D}-2\right) -\frac{x^2}{D} \frac{ \frac{5-D}{2} \; p^2}{x(1-x) p^2 + y \left(\frac{\Omega}{2} \right)^2  }\right] \equiv \delta \Pi^{V,3,1}_{xx,1} (i \Omega,0)+\delta \Pi^{V,3,2}_{xx,1} (i \Omega,0).
\end{eqnarray}
After the integral over the radial momentum variable $p$, one of the two entries in the final expression for $\delta \Pi^{V,3}_{xx,1} (i \Omega,0)$ reads as
\allowdisplaybreaks[4]
\begin{eqnarray}
&&\delta \Pi^{V,3,1}_{xx,1} (i \Omega,0) \nonumber \\
&=&\frac{N e^2}{32}   \frac{\Omega^{2D-4} 2^{8-2D} \pi^{1+D/2} \left( \frac{1}{D}-2\right) }{\Gamma(3/2) (4\pi)^{D/2} (2 \pi)^D \Gamma(D/2)} \int^1_0 dx \int^{1-x}_0 dy \sqrt{1-x-y} \bigg[ \left[x(1-x)-y \right]^{\frac{D-5}{2}} \; \frac{\pi \Gamma\left( \frac{5-D}{2} \right) }{\sin (\pi D)} \nonumber \\
&-& \Gamma(3-D) \Gamma \left( \frac{D-1}{2}\right) y^{D-3} \left[ x(1-x) \right]^{\frac{1-D}{2}} {_2}F_1 \left[1,\frac{D-1}{2},D-2,\frac{y}{x(1-x)} \right]  \bigg] =N e^2 \Omega^2 \frac{5}{192 \pi^3},
\end{eqnarray}
while the second term goes as
\begin{eqnarray}
&&\delta \Pi^{V,3,1}_{xx,1} (i \Omega,0) \nonumber \\
&=& \frac{N}{32} \Omega^{2D-4} \; \frac{2^{9-2D} \pi^{D/2} \Gamma\left( \frac{5-D}{2} \right) \left( \frac{5-D}{2D} \right)}{\Gamma(3/2) (4\pi)^{D/2} (2\pi)^D \Gamma\left( \frac{D}{2}\right)} \; \int^1_0 dx x^2 \int^{1-x}_0 dy (1-x-y)^{1/2}
\bigg[- \left[x(1-x)-y \right]^{\frac{D-7}{2}}  \nonumber \\
&\times& \frac{\pi}{\sin (\pi D)} + y^{D-3} \left[ x(1-x)\right]^{-\frac{D+1}{2}} \frac{\Gamma(3-D)\Gamma\left( \frac{D+1}{2}\right)}{\Gamma\left( \frac{7-D}{2}\right)}
{_2}F_1 \left[1,\frac{D+1}{2},D-2, \frac{y}{x(1-x)} \right] \bigg]= Ne^2 \Omega^2 \left[\frac{7 \zeta(3)-6 }{384 \pi^3} \right]. \nonumber \\
\end{eqnarray}
Therefore, the net contribution from $\delta \Pi^{V,3}_{xx,1} (i \Omega,0)$
\begin{equation}\label{eq:current-vertex-31-final}
\delta \Pi^{V,3}_{xx,1} (i \Omega,0)=\frac{Ne^2 \Omega^2}{\pi^3} \left[ \frac{5}{192}-\frac{6}{384} + \frac{7 \zeta(3)}{384} \right]
=N \frac{e^2 \Omega^2}{384 \pi^3} \left[ 4+ 7 \zeta(3) \right]
\end{equation}
is finite. Finally, the last term in the expression of $\delta \Pi^{V}_{xx} (i \Omega,0)$ is given by
\begin{eqnarray}\label{eq:current-vertex-2-final}
&&\delta \Pi^{V}_{xx,2} (i \Omega,0) \nonumber \\
&=&\pi e^2 \frac{N \Omega^2}{4} \frac{\Gamma\left( \frac{5}{2}\right)}{\Gamma\left( \frac{1}{2}\right)} \int^1_0 dx \int^{1-x}_0 dy \int \frac{d^D {\mathbf k}}{(2\pi)^D} \frac{ \left[1-x-y \right]^{-1/2}}{k \left[ k^2+(\Omega/2)^2 \right]} \; \int \frac{d^D {\mathbf p}}{(2\pi)^D} \; \frac{x \left[k^2-k^2_x \right]}{\left[ p^2 + x(1-x)k^2+ y (\Omega/2)^2 \right]^{5/2}} \nonumber \\
&=& \pi e^2 \frac{N \Omega^2}{4} \frac{ 2 \pi^{D/2}\left( 1-\frac{1}{D} \right)  \Gamma\left( \frac{5-D}{2}\right)}{ (2\pi)^D (4\pi)^{D/2} \Gamma\left( \frac{1}{2}\right) \Gamma\left( \frac{D}{2}\right)} \int^1_0 dx \int^{1-x}_0 dy \int^{\infty}_0 dk
\frac{k^d \; x \left[1-x-y \right]^{-1/2} \left[ x(1-x)\right]^{\frac{D-5}{2}}}{\left[ k^2+ \left(\frac{\Omega}{2} \right)^2 \right] \left[ k^2+ \frac{y}{x(1-x)}\left(\frac{\Omega}{2} \right)^2 \right]^{\frac{5-D}{2}}} \nonumber \\
&=& \frac{N e^2 \Omega^2}{8} \frac{(D-1) 2^{7-2D} \pi^{1+D/2} }{D (2\pi)^D (4\pi)^{D/2} \Gamma\left( \frac{1}{2}\right) \Gamma\left( \frac{D}{2}\right)} \; \int^1_0 dx \int^{1-x}_0 dy x \left[ 1-x-y\right]^{-1/2} \bigg\{ \left[x(1-x)-y \right]^{\frac{D-5}{2}} \frac{\pi \Gamma\left( \frac{5-D}{2}\right)}{\sin(\pi D)} \nonumber \\
&+& y^{D-2} \left[x(1-x) \right]^{-\frac{D+1}{2}} \Gamma(D-2) \Gamma\left( \frac{D+1}{2} \right) {_2}F_1 \left[ 1, \frac{D+1}{2}, D-1, \frac{y}{x(1-x)}\right]  \bigg\}
= \frac{N e^2 \Omega^2}{72 \pi^3} \left[ \frac{3}{2 \epsilon}\right].
\end{eqnarray}

Therefore the net divergent contribution arising from the vertex diagram is given by
\begin{equation}
\delta \Pi^{V}_{xx} (i \Omega,0)= -\frac{N e^2 \Omega^2}{72 \pi^3 v^2} \left[ \frac{2}{\epsilon^2} + \frac{15-4 \gamma_E + 4 \log(4\pi)}{2 \epsilon} \right].
\end{equation}
After accounting for the divergent piece coming from the self-energy diagram we obtain the net leading order correction to the polarization bubble due to the long-range tail of the Coulomb interaction to be
\begin{equation}\label{eq:current-Coulomb}
\delta \Pi_{xx} (i \Omega,0)= N \frac{e^2 \Omega^2}{72 \pi^3 v^2} \; \left[ \frac{1}{\epsilon^2} -\frac{1}{2 \epsilon} \left[1+2 \gamma_E -2\log(4\pi) \right] \right].
\end{equation}

\subsection{Density-density correlator}

Contribution to the density-density correlator from the self-energy diagram reads as
\begin{eqnarray}
\delta \Pi^{SE}_{00}(i\Omega,{\mathbf q})&=&-i^2 (2N)\int\frac{d\omega}{2\pi}\int\frac{d\omega'}{2\pi}\int \frac{d^D {\mathbf k}}{(2\pi)^D}\int \frac{d^D {\mathbf p}}{(2\pi)^D}{\rm \bf Tr}[G_f(i\omega,{\bf k})\, \sigma_0\, G_f(i\omega+i\Omega,{\bf k}+{\bf q}) \nonumber \\
&\times& \sigma_0\,G_f(i\omega,{\bf k})\,G_f(i\omega',{\bf p})\, V_C({\bf k}-{\bf p})],
\end{eqnarray}
or explicitly
\begin{eqnarray}
&& \delta \Pi^{SE}_{00}(i\Omega,{\mathbf q}) = - \frac{2 N}{v^{4}} \; {\rm \bf Tr} \bigg\{ \int \frac{d^D {\mathbf k}}{(2\pi)^D} \; \int \frac{d\omega}{2 \pi}  \; \frac{i \omega +  {\boldsymbol \sigma} \cdot {\mathbf k}}{\omega^2+ {\mathbf k}^2} \; \bigg[i^2 \int \frac{d^D {\mathbf p}}{(2\pi)^D} \; \int \frac{d\omega^\prime}{2 \pi} \frac{2 \pi e^2}{|{\mathbf k}-{\mathbf p}|^2} \;  \frac{i \omega^\prime +  {\boldsymbol \sigma} \cdot {\mathbf p}}{{\omega^\prime}^2+ {\mathbf p}^2} \bigg] \; \frac{i \omega +  {\boldsymbol \sigma} \cdot {\mathbf k}}{\omega^2+ {\mathbf k}^2} \nonumber \\
& \times & \frac{i (\omega+\Omega) + {\boldsymbol \sigma} \cdot \left( {\mathbf k} + {\mathbf  q} \right)}{(\omega+\Omega)^2+ \left( {\mathbf k} +{\mathbf q} \right)^2} \bigg\}
=-\frac{2N}{v^4} {\rm {Tr}} \bigg\{ \int \frac{d^D {\mathbf k}}{(2\pi)^D} \; \int \frac{d\omega}{2 \pi} \; \frac{i \omega +  {\boldsymbol \sigma} \cdot {\mathbf k}}{\omega^2+ {\mathbf k}^2} \;  \Sigma({\mathbf k}) \frac{i \omega +  {\boldsymbol \sigma} \cdot {\mathbf k}}{\omega^2+ {\mathbf k}^2} \; \frac{i (\omega+\Omega) + {\boldsymbol \sigma} \cdot \left( {\mathbf k} + {\mathbf q} \right)}{(\omega+\Omega)^2+ \left( {\mathbf k} +{\mathbf q} \right)^2}
\bigg\},\nonumber \\
\end{eqnarray}
with $\Sigma({\bf k})$ given by Eq.\ (\ref{eq:self-energy-1}).
After performing the trace algebra we arrive at the following compact expression for the self-energy contribution
\begin{eqnarray}
\delta \Pi^{SE}_{00}(i\Omega,{\mathbf q}) = -2 E(D) \; \int \frac{d^D {\mathbf k}}{(2\pi)^D} \; \int \frac{d\omega}{2 \pi} \: \: \frac{k^{D-3} \left[ -2 \omega(\omega+\Omega) k^2 + (k^2-\omega^2) \; {\mathbf k} \cdot ({\mathbf k}+{\mathbf q})   \right] }{(\omega^2 + k^2)^2 \; \left[ (\omega+\Omega)^2 + ({\mathbf k}+{\mathbf q})^2\right]}.
\end{eqnarray}
After expanding the above expression in powers of $q$ and retaining the terms only to the order $q^2$, we find $\delta \Pi^{SE}_{00}(i\Omega,{\mathbf q})=\delta \Pi^{SE,1}_{00}(i\Omega,{\mathbf q})+\delta \Pi^{SE,2}_{00}(i\Omega,{\mathbf q})$, where
\allowdisplaybreaks[4]
\begin{eqnarray}
&&\delta \Pi^{SE,1}_{00} (i\Omega,{\mathbf q}) \nonumber \\
&=& 4 E(D) \; \frac{q^2}{D} \; \int \frac{d^D {\mathbf k}}{(2\pi)^D} \; \int^{\infty}_{-\infty} \frac{d\omega}{2 \pi}  \frac{k^{D-1} (k^2-\omega^2)}{(\omega^2 + {\mathbf k}^2)^2 \; \left[ (\omega+\Omega)^2 + {\mathbf k}^2\right]^2} \nonumber \\
&=&  E(D) \; \frac{q^2}{D} \: \frac{2 \pi^{D/2}}{\Gamma\left( \frac{D}{2}\right) (2 \pi)^D} \: \int^{\infty}_0 dk \;  \frac{32 k^4-12 k^2 \Omega^2 -\Omega^4}{k^{5-2D}\left(4 k^2+\Omega^2 \right)} \nonumber \\
&=& E(D) q^2 \, \frac{2^{4-2D} \pi^{1+D/2}}{D \; \Gamma\left( \frac{D}{2}\right) (2 \pi)^D}  \left[ \left(5-7D+2 D^2 \right) \Omega^{2D-6} \mathrm{cosec} (\pi D) \right]=-\frac{N e^2 q^2}{36 \pi^3 v^2} \left[ \frac{1}{\epsilon^2} + \frac{1-\gamma_E + \log(4\pi)}{\epsilon}\right],
\end{eqnarray}
for $D=3-\epsilon$ and after taking $ q \to v q$. The remaining contribution from the self-energy diagram goes as
\allowdisplaybreaks[4]
\begin{eqnarray}
&&\delta \Pi^{SE,2}_{00}(i\Omega,{\mathbf q}) \nonumber \\
&=& -2 E(D) \; \frac{q^2}{D} \; \int \frac{d^D {\mathbf k}}{(2\pi)^D} \; \int^{\infty}_{-\infty} \frac{d\omega}{2 \pi} \; \frac{k^{D-1} \left[ -2 \omega(\omega+\Omega) + k^2-\omega^2 \right]}{(\omega^2 + {\mathbf k}^2)^2 \; \left[ (\omega+\Omega)^2 + {\mathbf k}^2\right]^2} \; \left[\frac{4 ({\mathbf k} \cdot {\mathbf q})^2}{(\omega+\Omega)^2 + k^2} -q^2\right] \nonumber \\
&=& 2 q^2 E(D) \frac{2 \pi^{D/2}}{\Gamma\left( \frac{D}{2}\right) (2 \pi)^D} \int^{\infty}_0 dk \:\:\: \frac{16 (D-5) k^4 + 24 k^2 \Omega^2 -(D-3)\Omega^4}{4 D k^{5-2D} \left[ \Omega^2 + 4 k^2\right]^3} \nonumber \\
&=& -q^2 \Omega^{2D-6} E(D) \; \frac{2^{3-2D} \pi^{1+ D/2}}{D \; \Gamma\left( \frac{D}{2}\right) (2 \pi)^D} (D-1)(2D-5) \mathrm{cosec} (\pi D)
= \frac{N e^2 q^2}{72 \pi^3 v^2} \left[ \frac{1}{\epsilon^2} + \frac{1-\gamma_E + \log(4\pi)}{\epsilon}\right].
\end{eqnarray}
Therefore, net self-energy correction reads as
\begin{equation}
\delta \Pi^{SE}_{00}(i\Omega,{\mathbf q}) =-\frac{N e^2 q^2}{72 \pi^3 v^2} \; \left[\frac{1}{\epsilon^2} +\frac{1-\gamma_E+\log(4\pi)}{\epsilon} \right].
\end{equation}

Next we turn our attention to the vertex diagram. The contribution from the vertex diagram in the presence of external frequency and momentum reads as
\begin{eqnarray}
\delta \Pi^{V}_{00} (i \Omega,{\mathbf q})&=&-i^2 N\int \frac{d^D {\mathbf k}}{(2\pi)^D} \int \frac{d^D {\mathbf p}}{(2\pi)^D} \int \frac{d\omega}{2 \pi}
\int \frac{d\omega^\prime}{2 \pi}{\rm \bf Tr}[G_f(i\omega,{\bf k})\,\sigma_0\,G_f(i\omega+i\Omega,{\bf k})\,G_f(i\omega+i\Omega,{\bf p}) \nonumber \\
&\times& \sigma_0\, G_f(i\omega,{\bf p})V_C({\bf k}-{\bf p})].
\end{eqnarray}
We now write this term as
\begin{eqnarray}
 \delta \Pi^{V}_{00}(i\Omega,{\mathbf q}) &=& \frac{N}{v^{4}} \; {\rm \bf Tr} \bigg\{ \int \frac{d^D {\mathbf k}}{(2\pi)^D} \; \int \frac{d^D {\mathbf p}}{(2\pi)^D} \; \frac{2 \pi e^2}{|{\mathbf k}-{\mathbf p}|^2} \; \left[ \int^{\infty}_{-\infty} \frac{d\omega}{2 \pi}  \;   \frac{i \left( \omega -\Omega \right) + {\boldsymbol \sigma} \cdot ({\mathbf k}-{\mathbf q})}{(\omega-\Omega)^2+ ({\mathbf k}-{\mathbf q})^2} \frac{i \omega +  {\boldsymbol \sigma} \cdot {\mathbf k}}{\omega^2+ {\mathbf k}^2} \right] \nonumber \\
&\times& \left[ \int^{\infty}_{-\infty} \frac{d\omega^\prime}{2 \pi} \; \frac{i \omega^\prime + {\boldsymbol \sigma} \cdot {\mathbf p}}{(\omega^\prime)^2+ {\mathbf p}^2} \frac{i \left( \omega^\prime -\Omega \right) +  {\boldsymbol \sigma} \cdot ({\mathbf p}-{\mathbf q})}{(\omega^\prime-\Omega)^2+ ({\mathbf k}-{\mathbf q})^2} \right]  \bigg\}.
\end{eqnarray}
The frequency integrals in the above expression can readily be performed following Eqs.~(\ref{I1_Vertex}) and (\ref{I2_Vertex}). Upon expanding four parameters $a,b,c,d$ defined through Eq.~(\ref{abcd}), up to the quadratic order in $q^2$ as shown in Eq.~(\ref{abcd_expansion}), we arrive at the following compact expression for the vertex function
\begin{eqnarray}
\delta \Pi^{V}_{00}(i\Omega,{\mathbf q}) = \frac{N}{2 v^{4}} \int \frac{d^D {\mathbf k}}{(2\pi)^D} \int \frac{d^D {\mathbf p}}{(2\pi)^D} \frac{2 \pi e^2}{|{\mathbf k}-{\mathbf p}|^2} \frac{ -\Omega^2 \left[ {\mathbf p} \left( {\mathbf p} \cdot {\mathbf q} \right) -p^2 {\mathbf q} \right] \cdot \left[ {\mathbf k} \left( {\mathbf k} \cdot {\mathbf q} \right) -k^2 {\mathbf q} \right]
+ 4 k^2 p^2 \left( {\mathbf p} \times {\mathbf q} \right) \cdot \left( {\mathbf k} \times {\mathbf q} \right)}{ k^3 p^3 \left[ \Omega^2 + 4 k^2 \right] \left[ \Omega^2 + 4 p^2 \right]}. \nonumber \\
\end{eqnarray}
The momentum integrals can be performed most efficiently by separating the above expression into two pieces yielding $\delta \Pi^{V}_{00}(i\Omega,{\mathbf q})=\left[ \delta \Pi^{V,1}_{00}(i\Omega,{\mathbf q})+\delta \Pi^{V,2}_{00}(i\Omega,{\mathbf q}) \right]/v^4$. Now we present evaluation of each term. The two terms are of the form
\begin{eqnarray}
\delta \Pi^{V,1}_{00}(i\Omega,{\mathbf q}) &=& N \int \frac{d^D {\mathbf k}}{(2\pi)^D} \frac{\left[ {\mathbf k} \left( {\mathbf k} \cdot {\mathbf q} \right) -k^2 {\mathbf q} \right]}{k^3\left[ \Omega^2 + 4 k^2 \right]} \cdot {\mathbf I}_1 ({\mathbf q}, {\mathbf k},\Omega), \nonumber \\
\delta \Pi^{V,2}_{00}(i\Omega,{\mathbf q}) &=& 2N \int \frac{d^D {\mathbf k}}{(2\pi)^D} \int \frac{d^D {\mathbf p}}{(2\pi)^D} \; \frac{2 \pi e^2}{|{\mathbf k}-{\mathbf p}|^2} \; \frac{\left( {\mathbf p} \times {\mathbf q} \right) \cdot \left( {\mathbf k} \times {\mathbf q} \right)}{k p \left[ \Omega^2 + 4 k^2 \right] \left[ \Omega^2 + 4 p^2 \right]}.
\end{eqnarray}
The quantity ${\mathbf I}_1 ({\mathbf q}, {\mathbf k},\Omega)$ appearing in the expression of $\delta \Pi^{V,1}_{00}(i\Omega,{\mathbf q})$ reads as
\allowdisplaybreaks[4]
\begin{eqnarray}
&&{\mathbf I}_1 ({\mathbf q}, {\mathbf k},\Omega) = -\frac{\Omega^2}{8} \int \frac{d^D {\mathbf p}}{(2\pi)^D} \; \frac{2 \pi e^2}{|{\mathbf k}-{\mathbf p}|^2} \; \frac{\left[ {\mathbf p} \left( {\mathbf p} \cdot {\mathbf q} \right) -p^2 {\mathbf q} \right]}{p^3 \left[ \Omega^2 + 4 p^2 \right]} \nonumber \\
&=& -\frac{2 \pi e^2 \Omega^2}{8} \frac{\Gamma\left( \frac{7}{2} \right)}{\Gamma\left( \frac{3}{2} \right)} \; \int^1_0 dx \int^{1-x}_0 dy \sqrt{1-x-y} \; \int \frac{d^D {\mathbf p}}{(2\pi)^D} \; \frac{ {\mathbf p} \left({\mathbf p} \cdot {\mathbf q} \right) + x^2 {\mathbf k} \left({\mathbf k} \cdot {\mathbf q} \right) -p^2 {\mathbf q} -x^2 k^2 {\mathbf q} }{\left[ p^2 + x(1-x) k^2 + y \left( \Omega/2 \right)^2 \right]^{7/2}} \nonumber \\
&=& -\frac{2 \pi e^2 \Omega^2}{8 \Gamma\left( \frac{3}{2} \right) (4\pi)^{D/2}} \int^1_0 dx \int^{1-x}_0 dy \sqrt{1-x-y} \left[
\frac{(1-D) \Gamma\left( \frac{5-D}{2}\right) {\mathbf q}}{2 \left[ x(1-x) k^2 + y \left( \Omega/2 \right)^2  \right]^{\frac{5-D}{2}}}
+ \frac{x^2 \left[ {\mathbf k} \left( {\mathbf k} \cdot {\mathbf q} \right) -k^2 {\mathbf q} \right] \Gamma \left( \frac{7-D}{2}\right)}{\left[ x(1-x) k^2 + y \left( \Omega/2 \right)^2  \right]^{\frac{7-D}{2}}}
\right]. \nonumber \\
\end{eqnarray}
After some algebraic simplifications we arrive at the following expression for
\begin{eqnarray}
\delta \Pi^{V,1}_{00}(i\Omega,{\mathbf q})= N q^2 \Omega^2 V(D) \int^1_0 dx \int^{1-x}_0 dy \frac{\sqrt{1-x-y}}{\left[ x(1-x) \right]^{\frac{7-D}{2}}} \int^{\infty}_0 dk \; \frac{A(x, D) \; k^D + B(y,D) \Omega^2 \;  k^{D-2}}{\left[ k^2+ \left( \frac{\Omega}{2} \right)^2 \right] \left[ k^2+ \frac{y}{x(1-x)}\left( \frac{\Omega}{2} \right)^2 \right]^{\frac{7-D}{2}}},
\end{eqnarray}
where $x$ and $y$ are Feynman parameters and
\begin{eqnarray}
V(D)= \frac{\pi e^2 \left(1- \frac{1}{D}\right)\pi^{1+\frac{D}{2}}}{16 \Gamma \left( \frac{D}{2}\right) (2 \pi)^D (4\pi)^{\frac{D}{2}}}, \:
A(x,D)=\frac{x(1-x)\frac{D-1}{2}\Gamma\left(\frac{5-D}{2} \right) -2 x^2 \Gamma\left(\frac{7-D}{2} \right)}{2}, \:
B(y,D) = \frac{(D-1) \Gamma\left(\frac{5-D}{2} \right) }{8}.
\end{eqnarray}
After some lengthy algebra it can be shown that $\delta \Pi^{V,1}_{00}(i\Omega,{\mathbf q}) = {\mathcal O}(\epsilon)$ and thus does not contribute to the conductivity. On the other hand, after algebraic manipulation the second term in the expression of $\delta \Pi^{V}_{00}(i\Omega,{\mathbf q})$ becomes
\begin{eqnarray}
&&\delta \Pi^{V,2}_{00}(i\Omega,{\mathbf q})=N \pi e^2 \frac{\Gamma\left( \frac{5}{2}\right)}{\Gamma\left( \frac{1}{2}\right)}\int^1_0 dx  \int^{1-x}_0 \frac{x \; dy}{ \sqrt{1-x-y}} \int \frac{d^D {\mathbf k}}{(2\pi)^D}  \int \frac{d^D {\mathbf p}}{(2\pi)^D} \; \frac{\left( {\mathbf k} \times {\mathbf q} \right)^2}{\left[ p^2 + x(1-x) k^2 + y (\Omega/2)^2 \right]^{5/2}} \nonumber \\
&=& N e^2 q^2 \frac{\left( 1-\frac{1}{D}\right)\Gamma\left( \frac{5-D}{2}\right) \pi^{1+ D/2}}{2 \Gamma\left( \frac{1}{2}\right) \Gamma\left( \frac{D}{2}\right) (2\pi)^D (4 \pi)^{D/2}} \int^1_0 dx \int^{1-x}_0 dy
\int^{\infty}_0 dk \frac{ x \left[ x(1-x) \right]^{\frac{D-5}{2}} \left[ 1-x-y\right]^{-1/2}}{\left[ k^2+ \left( \frac{\Omega}{2} \right)^2 \right] \left[ k^2+ \frac{y}{x(1-x)}\left( \frac{\Omega}{2} \right)^2 \right]^{\frac{5-D}{2}}} \; k^D.
\end{eqnarray}
After a lengthy calculation and at the end setting $D=3-\epsilon$ and $q \to v q$ we arrive at the final expression for the vertex correction to the density-density correlator coming from the Coulomb interaction
\begin{eqnarray}
\delta \Pi^{V}_{00}(i\Omega,{\mathbf q})= \frac{N e^2 q^2}{72 \pi^3 v^2} \; \left[\frac{3}{2} \times \frac{1}{\epsilon} \right] + {\mathcal O} (1).
\end{eqnarray}
Therefore, after accounting for the contribution from self-energy diagram we obtain the net correction to the polarization tensor due to the long-range tail of the Coulomb interaction from density-density correlation to be
\begin{equation}\label{eq:density-Coulomb}
\delta \Pi_{00} (i \Omega, {\mathbf q}) = \delta \Pi^{SE}_{00} (i \Omega, {\mathbf q}) + \delta \Pi^{V}_{00} (i \Omega, {\mathbf q})
=- \frac{N e^2 q^2}{72 \pi^3 v^2} \; \left[ \frac{1}{\epsilon^2} - \frac{1}{2 \epsilon} \left[ 1+ 2 \gamma_E -2 \log(4\pi) \right] \right]
\equiv -\frac{N e^2 q^2}{72 \pi^3 v^2} \; \left[ \frac{1}{\epsilon^2} + \frac{b}{\epsilon} \right],
\end{equation}
where $b=-\left[ 1+ 2 \gamma_E -2 \log(4\pi) \right]/2 \approx 1.454$.
\\
 We point our here that the results for the Coulomb correction to the current-current correlator in Eq. (\ref{eq:current-Coulomb}) and to the density-density correlator in Eq. (\ref{eq:density-Coulomb}) are consistent with the charge conservation, Eq. (\ref{CC-2}), and therefore
 dimensional regularization employed manifestly preserves gauge invariance of the theory.


\subsection{Correction to optical conductivity}~\label{OC:final}

Upon obtaining the same expression for the polarization bubble from both current-current and density-density correlators we can proceed with the computation of the correction to the OC due to the long-range piece of the Coulomb interaction. Following the steps highlighted in Sec.~\ref{OC:noninteracting}, we find
\begin{equation}
\sigma_{jj} (\Omega)= \sigma_0 (\Omega) \; \left[ 1- \frac{\alpha}{6\pi}\left\{ b-2 \log\left( \frac{E_\Lambda}{\Omega} \right) \right\} \right],
\end{equation}
where $\alpha=e^2/(\hbar v)$ is the fine structure constant in the Weyl medium. Here $E_\Lambda=2 v \Lambda$, where $\Lambda$ is the ultraviolet momentum cut-off and therefore $E_\Lambda$ is the band width of a Weyl semimetal (WSM) in linearized approximation. If we also account for the correction to the OC due to the short-range piece of the density-density interaction the OC of an interacting Weyl liquid (to the leading order in coupling constants) is given by
\begin{equation}~\label{OC:corrected}
\sigma_{jj} (\Omega)= \sigma_0 (\Omega) \; \left[ 1- \frac{\alpha}{6\pi}\left\{ b-2 \log\left( \frac{E_\Lambda}{\Omega} \right) \right\}
+ \left( \frac{g_0 \Omega^2}{24 \pi^2 v^3}\right) \;\left\{ a- 2 \log\left( \frac{E_\Lambda}{ \Omega}\right) \right\} \right],
\end{equation}
with $a \approx 3.62069$ and $b \approx 1.454$. Recall $\sigma_0 (\Omega)$ is the conductivity of a $N$-flavored non-interacting Weyl semimetal, given by $\sigma_0(\Omega)=N e^2_0/(12 h v)$, where $N$ is the number of Weyl points in the Brillouin zone.
\\

We realize that coupling constants appearing in the above expression for the optical conductivity are renormalized ones. Therefore, two dimensionless couplings, namely $\alpha$ and $g=g_0 \Omega^2/v^3$ in the above expression needs to be substituted by their scale dependent strengths, which we can readily obtained from the renormalization group flow equations of these two couplings appearing respectively in Eq.~(\ref{beta:alpha}) and Eq.~(\ref{beta:g0}). After expressing $\beta_{x}$ as $dx/dl$, where $x=\alpha, \hat{g}_0$ and $l$ is the logarithm of the renormalization scale, and therefore $l=\log \left( E_\Lambda/\Omega \right)$, we obtain the running couplings
\begin{equation}
\alpha(\Omega)= \frac{\alpha_0}{1+ \alpha_0 \frac{N+1}{3 \pi} \log \left( \frac{E_\Lambda}{\Omega} \right) } \approx \frac{1}{\frac{N+1}{3 \pi} \log \left( \frac{E_\Lambda}{\Omega} \right)}, \quad
g(\Omega)=\hat{g}_0 \; \left( \frac{\Omega}{E_\Lambda} \right)^2,
\end{equation}
where the quantities with subscript ``$0$" denote their bare strength. Upon substituting these running couplings back in Eq.~(\ref{OC:corrected}), we find
\begin{equation}
\sigma_{jj} (\Omega)= \sigma_0 (\Omega) \; \left[1+\frac{1}{N+1}-\frac{b}{2 (N+1) \log\left(\frac{E_\Lambda}{\Omega}\right)}-\frac{\hat{g}_0\Omega^2}{12\pi^2 E^2_\Lambda}
\left\{ \log\left(\frac{E_\Lambda}{\Omega}\right)-\frac{a}{2}\right\} \right],
\end{equation}
which matches with Eq.~(13) of the main text, and ultimately simplifies to Eq.~(1) from the main text for $\Omega \ll E_\Lambda$.


\section{Alternative computation of current-current correlator with long-range Coulomb interaction}~\label{OC:alternative}

We here present an alternative route to compute the current-current correlator in the presence of long-range Coulomb interaction. In the previous calculation we have chosen $l=m=x$ at the fermion-current vertex and performed the analysis. Alternatively, we can sum over this contribution for all spatial components of the current operator, and then devide the final expression by $D$ (spatial dimensionality) to obtain the contribution of each diagram/term to longitudinal conductivity.
\\

If we sum over the contribution from the self-energy diagram from all spatial components of current operators, its contribution to conductivity goes as (after performing the trace algebra)
\begin{eqnarray}
\delta \Pi^{SE}_{lm}(i\Omega,0) &=& \frac{4 N E(D)}{D}\;\delta_{lm} \; \int \frac{d^D {\mathbf k}}{(2\pi)^D}  \int^{\infty}_{-\infty} \frac{d\omega}{2 \pi} \: \frac{ \left[ -2D k^2 \omega(\omega+\Omega) -(2-D)\left( \omega^2 k^2 - k^4\right) \right]}{k^{3-D} \left[ \omega^2+k^2 \right]^2 \left[(\omega+\Omega)^2+ k^2 \right]} \nonumber \\
&=&\frac{N e^2 \Omega^2}{72 \pi^3 v^2} \; \left[ \frac{3}{\epsilon^2} + \frac{1}{\epsilon} \left[ 7-3 \gamma_E + 3 \log(4 \pi)\right]\right]\;\delta_{lm},
\end{eqnarray}
which is identical to the result obtained in Eq.\ (\ref{eq:current-selfenergy-final}).
Since the frequency integral produces a lengthy expression, we here do not wish to present all the intermediate steps, which, however, follow those in Eq.\ (\ref{eq:current-selfenergy-final}).
\\

We can proceed with the same strategy for the vertex diagram by computing the terms in Eqs.\ (\ref{eq:current-vertex-11})-(\ref{eq:current-vertex-2}). In this framework $\delta \Pi^{V,2}_{lm,1} (i \Omega,0)$ remains unchanged, while three other components appearing in $\delta \Pi^{V}_{lm,1} (i \Omega,0)$ read as
\allowdisplaybreaks[4]
\begin{eqnarray}
\delta \Pi^{V,1}_{lm,1} (i \Omega,0) &=& \frac{N\Omega^2}{4 D} \delta_{lm}\int \frac{d^D {\mathbf k}}{(2\pi)^D} \int \frac{d^D {\mathbf p}}{(2\pi)^D} \frac{2 \pi e^2}{|{\mathbf k}-{\mathbf p}|^2} \; \frac{ \left( {\mathbf k} \cdot {\mathbf p} \right)^2 +(D-2) k^2 p^2 }{k^3 p^3 \left[ k^2 + (\Omega/2)^2 \right]}, \\
\delta \Pi^{V,3}_{lm,1} (i \Omega,0) &=& -\frac{N\Omega^4}{32 D} \delta_{lm}\int \frac{d^D {\mathbf k}}{(2\pi)^D} \int \frac{d^D {\mathbf p}}{(2\pi)^D} \frac{2 \pi e^2}{|{\mathbf k}-{\mathbf p}|^2}  \frac{ \left( {\mathbf k} \cdot {\mathbf p} \right)^2 - 2 p^2 k^2}{k^3 p^3 \left[ k^2 + (\Omega/2)^2 \right] \left[ p^2 + (\Omega/2)^2 \right]}, \\
\delta \Pi^{V}_{lm,2} (i \Omega,0) &=& \frac{N\Omega^2}{8 D} (D-1) \delta_{lm}\int \frac{d^D {\mathbf k}}{(2\pi)^D} \int \frac{d^D {\mathbf p}}{(2\pi)^D} \frac{2 \pi e^2}{|{\mathbf k}-{\mathbf p}|^2} \frac{{\mathbf k} \cdot {\mathbf p}}{k p \left[ k^2 + (\Omega/2)^2 \right] \left[ p^2 + (\Omega/2)^2 \right]}.
\end{eqnarray}
Now we present some key steps of the computation of each term and display the final result.
\\

After some algebraic simplification $\delta \Pi^{V,1}_{lm,1} (i \Omega,0)$ reads as
\begin{eqnarray}
\delta \Pi^{V,1}_{lm,1} (i \Omega,0) = \frac{N \Omega^2}{4D} \: \left[ \frac{1}{4}I_1 + \frac{1}{4} I_2 + \left(D-\frac{2}{3} \right) I_3 \right]\;\delta_{lm}.
\end{eqnarray}
The integral $I_1$ in the above expression goes as
\begin{eqnarray}
I_1 &=& \int \frac{d^D {\mathbf k}}{(2\pi)^D} \int \frac{d^D {\mathbf p}}{(2\pi)^D}  \frac{k}{ p^3 |{\mathbf k}-{\mathbf p} |^2 \left[ k^2+(\Omega/2)^2 \right]}
=\frac{2 \pi^{D/2}\Gamma\left( \frac{5-D}{2}\right) \Gamma\left( \frac{D-3}{2}\right) \Gamma\left( \frac{D}{2}-1\right)}{(4\pi)^{D/2} (2\pi)^{D} \Gamma\left( \frac{3}{2}\right) \Gamma\left( \frac{D}{2}\right) \Gamma\left( D-\frac{5}{2}\right) } \int^\infty_0 dk \frac{k^{2D-5}}{k^2+(\Omega/2)^2} \nonumber \\
&=& \frac{2 \pi^{D/2}\Gamma\left( \frac{5-D}{2}\right) \Gamma\left( \frac{D-3}{2}\right) \Gamma\left( \frac{D}{2}-1\right)}{(4\pi)^{D/2} (2\pi)^{D} \Gamma\left( \frac{3}{2}\right) \Gamma\left( \frac{D}{2}\right) \Gamma\left( D-\frac{5}{2}\right) } \: \left[ \Omega^{2D-6} \; 2^{5-2D}\; \pi \; \mathrm{cosec}(\pi D) \right].
\end{eqnarray}
The quantity in the straight bracket comes from the integral over radial momentum variable $k$. The following integral identity will be extremely useful to compute $I_2$ and $I_3$
\begin{equation}
\int \frac{d^D {\mathbf p}}{(2\pi)^D}  \: \frac{1}{|{\mathbf k}-{\mathbf p} |^2 \left( |{\mathbf p}|^2 \right)^a}
= \frac{\Gamma \left( 1+a-\frac{D}{2} \right) \Gamma \left(\frac{D}{2}-1 \right) \Gamma \left(\frac{D}{2}-a \right)}{ \Gamma(a) \Gamma(D-1-a)} \: \left(k^2 \right)^{-1-a-\frac{D}{2}}.
\end{equation}
With the help of the above integral identity, the second term is given by
\begin{eqnarray}
I_2&=& \int \frac{d^D {\mathbf k}}{(2\pi)^D} \int \frac{d^D {\mathbf p}}{(2\pi)^D}  \frac{p}{ k^3 |{\mathbf k}-{\mathbf p} |^2 \left[ k^2+(\Omega/2)^2 \right]}
=\frac{2 \pi^{D/2}\Gamma\left( \frac{1-D}{2}\right) \Gamma\left( \frac{D+1}{2}\right) \Gamma\left( \frac{D}{2}-1\right)}{(4 \pi)^{D/2} (2 \pi)^D \Gamma\left( \frac{D}{2}\right) \Gamma\left( -\frac{1}{2}\right) \Gamma\left( D-\frac{1}{2}\right) }  \int^\infty_0 dk \:  \frac{k^{2D-5}}{k^2+(\Omega/2)^2} \nonumber \\
&=& \frac{2 \pi^{D/2}\Gamma\left( \frac{1-D}{2}\right) \Gamma\left( \frac{D+1}{2}\right) \Gamma\left( \frac{D}{2}-1\right)}{(4 \pi)^{D/2} (2 \pi)^D \Gamma\left( \frac{D}{2}\right) \Gamma\left( -\frac{1}{2}\right) \Gamma\left( D-\frac{1}{2}\right) }
\: \left[ \Omega^{2D-6} \; 2^{5-2D}\; \pi \; \mathrm{cosec}(\pi D) \right].
\end{eqnarray}
The last term in the expression of $\delta \Pi^{V,1}_{lm,1} (i \Omega,0)$ assumes the form
\begin{eqnarray}
I_3 &=& \int \frac{d^D {\mathbf k}}{(2\pi)^D} \int \frac{d^D {\mathbf p}}{(2\pi)^D} \frac{1}{ k p |{\mathbf k}-{\mathbf p} |^2 \left[ k^2+(\Omega/2)^2 \right]}
= \frac{2 \pi^{D/2}\Gamma\left( \frac{3-D}{2}\right) \Gamma\left( \frac{D-1}{2}\right) \Gamma\left( \frac{D}{2}-1\right)}{(4 \pi)^{D/2} (2 \pi)^D \Gamma\left( \frac{D}{2}\right) \Gamma\left( \frac{1}{2}\right) \Gamma\left( D-\frac{3}{2}\right) } \: \int^\infty_0 dk \:  \frac{k^{2D-5}}{k^2+(\Omega/2)^2} \nonumber \\
&=& \frac{2 \pi^{D/2}\Gamma\left( \frac{3-D}{2}\right) \Gamma\left( \frac{D-1}{2}\right) \Gamma\left( \frac{D}{2}-1\right)}{(4 \pi)^{D/2} (2 \pi)^D \Gamma\left( \frac{D}{2}\right) \Gamma\left( \frac{1}{2}\right) \Gamma\left( D-\frac{3}{2}\right) } \: \left[ \Omega^{2D-6} \; 2^{5-2D}\; \pi \; \mathrm{cosec}(\pi D) \right].
\end{eqnarray}
Now combining the contributions from $I_1$, $I_2$ and $I_3$, we obtain
\begin{equation}
\delta \Pi^{V,1}_{lm,1} (i \Omega,0)= \frac{Ne^2 \Omega^2}{72 \pi^3 } \left[ \frac{2}{\epsilon^2} + \frac{6-2\gamma_E +2 \log(4\pi)}{\epsilon} \right]\;\delta_{lm},
\end{equation}
for $D=3-\epsilon$, which is in agreement with the result in  Eq.\ (\ref{eq:current-vertex-11-final}).
\\

Now we proceed with the computation of $\delta \Pi^{V,3}_{lm,1} (i \Omega,0)$. After some simple algebraic manipulation we find
\begin{eqnarray}
\delta \Pi^{V,3}_{lm,1} (i \Omega,0)= - 2\pi e^2 \; \frac{N\Omega^4}{32D} \left[K_1-2 K_2 \right]\;\delta_{lm},
\end{eqnarray}
where $K_1=\frac{1}{2} \left( J_1 + J_2 + J_3 \right)$, with
\begin{eqnarray}
J_1=-\int \frac{d^D {\mathbf k}}{(2\pi)^D} \int \frac{d^D {\mathbf p}}{(2\pi)^D} \frac{1}{k^3 p^3 \left[ k^2+(\Omega/2)^2 \right] \left[ p^2+(\Omega/2)^2 \right] }= \Omega^{2D-8} \; (16 \pi)^{2-D} \; \frac{\sec^2 \left( \frac{\pi D}{2}\right)}{\left[ \Gamma\left( \frac{D}{2} \right) \right]^2}.
\end{eqnarray}
The second term in $K_1$ reads as
\allowdisplaybreaks[4]
\begin{eqnarray}
J_2 &=& \int \frac{d^D {\mathbf k}}{(2\pi)^D}  \frac{k}{\left[ k^2+(\Omega/2)^2 \right]} \int \frac{d^D {\mathbf p}}{(2\pi)^D}  \frac{1}{p^3 |{\mathbf k}-{\mathbf p}|^2 \left[ p^2+(\Omega/2)^2 \right] } \nonumber \\
&=& \frac{\Gamma\left( \frac{7-D}{2}\right)}{(4\pi)^{D/2} \Gamma\left( \frac{3}{2}\right)}  \frac{2 \pi^{D/2}}{(2\pi)^D \Gamma\left( D/2\right)} \int^1_0 dx \int^{1-x}_0 dy \int^\infty_0 dk  \; \frac{k^D \: \sqrt{1-x-y} \: \left[ x(1-x)\right]^{\frac{D-7}{2}}}{ \left[ k^2+(\Omega/2)^2 \right] \left[ k^2 + \frac{y}{x(1-x)} (\Omega/2)^2 \right]^{\frac{D-7}{2}}}.
\end{eqnarray}
Further analysis of $J_2$ produces extremely lengthy expression and we perform the analysis in mathematica. The last term in $K_1$ goes as
\begin{eqnarray}
J_3= \int \frac{d^D {\mathbf k}}{(2\pi)^D} \int \frac{d^D {\mathbf p}}{(2\pi)^D} \frac{1}{k p |{\mathbf k}-{\mathbf p}|^2 \left[ k^2+(\Omega/2)^2 \right] \left[ p^2+(\Omega/2)^2 \right]} = \left[ N e^2 \Omega^2 \: \frac{7 \zeta(3)}{128 \pi^3} \right] \times \frac{32}{2 \pi N \Omega^2 e^2},
\end{eqnarray}
and $K_2= J_3$. Upon collecting all these contribution we arrive at the final expression for $\delta \Pi^{V,3}_{lm,1} (i \Omega,0)$, given by
\begin{eqnarray}
\delta \Pi^{V,3}_{lm,1} (i \Omega,0) &=& -2\pi e^2\frac{N\Omega^4}{32D} \left[ \frac{1}{2} \left(J_1+J_2 \right)-\frac{3}{2} J_3 \right]\delta_{lm}
\approx \frac{N e^2 \Omega^2}{\pi^3} \left[ -\frac{0.000531776}{96} \pi^4 + \frac{7 \zeta(3)}{256} \right] \;\delta_{lm}\nonumber \\
&\approx& \frac{N e^2 \Omega^2}{\pi^3} \left(0.0323292 \right)\;\delta_{lm}
= \frac{N e^2 \Omega^2}{384 \pi^3} \; \left[ 4+ 7 \zeta(3) \right]\;\delta_{lm},
\end{eqnarray}
in agreement with the result in Eq.\ (\ref{eq:current-vertex-31-final}).

Finally we come to the computation of $\delta \Pi^{V}_{lm,2} (i \Omega,0)$, which after some algebraic simplification can be expressed as
\begin{equation}
\delta \Pi^{V}_{lm,2} (i \Omega,0)= -\pi e^2 \frac{N \Omega^2}{8} \left( 1-\frac{1}{D} \right) \: \left[ \left( L_1 \right)^2 -2 L_2 + 2 L_3\right]\;\delta_{lm},
\end{equation}
where
\begin{eqnarray}
L_1=\int \frac{d^D {\mathbf k}}{(2\pi)^D} \; \frac{1}{k \left[ k^2 + (\Omega/2)^2 \right]}
= \frac{2 \pi^{D/2}}{\Gamma\left( \frac{D}{2}\right) (2 \pi)^D } \: \left[ - \pi \; 2^{2-D} \; \Omega^{D-3} \; \mathrm{sec}\left( \frac{\pi D}{2}\right) \right].
\end{eqnarray}
The second entry in the expression of $\delta \Pi^{V}_{lm,2} (i \Omega,0)$ goes as
\begin{eqnarray}
L_2 &=& \int \frac{d^D {\mathbf k}}{(2\pi)^D} \int \frac{d^D {\mathbf p}}{(2\pi)^D} \; \frac{1}{k p |{\mathbf k}-{\mathbf p}|^2 \left[ p^2 + (\Omega/2)^2 \right]} \nonumber \\
&=& \frac{\Gamma \left( \frac{5-D}{2} \right) }{(4\pi)^{D/2} \Gamma\left( \frac{1}{2}\right) } \frac{2 \pi^{D/2}}{(2 \pi)^D  \Gamma\left( \frac{D}{2}\right) } \int^1_0 dx \left[ x(1-x)\right]^{\frac{D-5}{2}} \int^{1-x}_0 dy
 \int^{\infty}_0 dk \frac{k^{D-2} \; \left[1-x-y \right]^{-1/2}}{\left[ k^2 + \frac{y}{x(1-x)} \left( \frac{\Omega}{2} \right)^2 \right]^{\frac{5-D}{2}}} \nonumber \\
&=& \frac{2 \pi^{D/2} \; \Gamma(3-D) \Gamma\left( \frac{D-1}{2}\right)}{(2 \pi)^D (4\pi)^{D/2} \Gamma\left( \frac{D}{2}\right) \Gamma\left( \frac{1}{2}\right) } 2^{5-2D} \left(\Omega^2\right)^{D-3} \int^1_0 dx \left[x(1-x) \right]^{\frac{1-D}{2}} \int ^{1-x}_0 dy y^{D-3} \; \left[ 1-x-y \right]^{-1/2} \nonumber \\
&=& \frac{2 \pi^{D/2} \; \Gamma(3-D) \Gamma\left( \frac{D-1}{2}\right)}{(2 \pi)^D (4\pi)^{D/2} \Gamma\left( \frac{D}{2}\right) \Gamma\left( \frac{1}{2}\right) } 2^{5-2D} \left(\Omega^2\right)^{D-3} \; \frac{\Gamma(D-2) \Gamma\left( \frac{3-D}{2}\right) \Gamma\left( \frac{D}{2}-1\right)}{\Gamma\left( D-\frac{3}{2}\right)}.
\end{eqnarray}
The last entry in the expression of $\delta \Pi^{V}_{lm,2} (i \Omega,0)$ is given by
\begin{eqnarray}
L_3 &=& \left( \frac{\Omega}{2} \right)^2 \int \frac{d^D {\mathbf k}}{(2\pi)^D} \int \frac{d^D {\mathbf p}}{(2\pi)^D}
\frac{1}{k p |{\mathbf k}-{\mathbf p}|^2 \left[ p^2 + (\Omega/2)^2 \right] \left[ k^2 + (\Omega/2)^2 \right]} \nonumber \\
&=& \left( \frac{\Omega}{2} \right)^2 \frac{\Gamma \left( \frac{5-D}{2} \right) }{(4\pi)^{D/2} \Gamma\left( \frac{1}{2}\right) } \frac{2 \pi^{D/2}}{(2 \pi)^D  \Gamma\left( \frac{D}{2}\right) } \int^1_0 dx  \int^{1-x}_0 dy \int^{\infty}_0 dk
\frac{k^{D-2} \; \left[ x(1-x)\right]^{\frac{D-5}{2}}  \left[1-x-y \right]^{-1/2}}{\left[ k^2 + \left(\frac{\Omega}{2} \right)^2 \right] \left[ k^2 + \frac{y}{x(1-x)} \left( \frac{\Omega}{2} \right)^2 \right]^{\frac{5-D}{2}} } \nonumber \\
&=& \frac{1}{32 \pi^4} \int^1_0 dx \int^{1-x}_0 dy \; \frac{\left[1-x-y \right]^{-1/2}}{x(1-x)} \; \frac{\log\left[ \frac{y}{x(1-x)}\right]}{\frac{y}{x(1-x)}-1}
=\frac{7 \zeta(3)}{32 \pi^4}.
\end{eqnarray}
Now collecting the contributions from $L_1$, $L_2$ and $L_3$ we obtain the divergent piece of $\delta \Pi^{V}_{lm,2} (i \Omega,0)$ to be
\begin{eqnarray}
\delta \Pi^{V}_{lm,2} (i \Omega,0) =\frac{N e^2 \Omega^2}{72 \pi^3} \; \left[ \frac{3}{2 \epsilon}\right]\;\delta_{lm},
\end{eqnarray}
in agreement with the result in Eq.\ (\ref{eq:current-vertex-2-final}).
Thus in this alternative approach to compute the current-current correlator we obtain identical results for each and every contribution to  both self-energy and vertex diagram.

\section{Kramers-Kronig relations and dielectric constant}~\label{append:dielectric}

Finally, we present the computation of the imaginary part of the optical conductivity [$\Im (\sigma)$], which is tied with the real part of the dielectric constant [$\varepsilon(\Omega)$] according to
\begin{equation}~\label{dielectric:Def}
\varepsilon(\Omega)=1-\frac{4\pi}{\Omega} \Im (\sigma),
\end{equation}
from its real component [$\Re(\sigma)$] by applying the Kramers-Kronig relation
\begin{equation}
\Im (\sigma) = -\frac{2 \pi}{\Omega} \; {\mathcal P} \; \int^{E_\Lambda}_0 \; d\Omega^\prime \: \frac{\Re(\sigma)}{ {\Omega^\prime}^2 -\Omega^2}.
\end{equation}
In the above expression ${\mathcal P}$ denotes the \emph{principle value} of the integral. Since we are interested in the regime $\Omega \ll E_\Lambda$ so that signature of Weyl fermions are prominent, we take the simplified expression for the real part of the optical conductivity after accounting for the leading correction due to Coulomb interaction, given by [Eq.~(1) of main text]
\begin{equation}
\Re(\sigma)= \sigma_0 (\Omega) \left[ 1+ \frac{1}{N+1}\right].
\end{equation}
The corresponding imaginary part of the optical conductivity is then given by
\begin{equation}
\Im(\sigma)=- \frac{e^2_0}{h} \; \frac{N \Omega}{6 \pi v} \left[ 1+ \frac{1}{N+1}\right] \; \log\left( \frac{E_\Lambda}{\Omega} \right).
\end{equation}
The above expression in conjunction with the definition of the real part of the dielectric constant [see Eq.~(\ref{dielectric:Def})] leads to
\begin{equation}
\varepsilon(\Omega)=1+\frac{2N e^2}{3 h v} \left[1+\frac{1}{N+1} \right] \; \log\left( \frac{E_\Lambda}{\Omega}\right),
\end{equation}
in agreement with Eq.~(14) of the main text.

\twocolumngrid


\begin{thebibliography}{}

\bibitem{basov2005} D.~ N.~ Basov and T.~ Timusk, Rev. Mod. Phys. {\bf 77}, 722 (2005).

\bibitem{Degiorgi1999} L.~ Degiorgi, Rev. Mod. Phys. {\bf 71}, 687 (1999).

\bibitem{Haule2011} D. N. Basov, R. D. Averitt, D. van der Marel, M. Dressel, and K. Haule, Rev. Mod. Phys. {\bf 83}, 471 (2011).

\bibitem{Si2009} Q.~ Si, Nat. Phys. {\bf 5}, 629 (2009).

\bibitem{Degiorgi2011} L.~ Degiorgi, New. J. Phys. {\bf 13}, 023011 (2011).


\bibitem{graphene:OC-1} R. R. Nair, P. Blake, A. N. Grigorenko, K. S. Novoselov, T. J. Booth, T. Stauber, N. M. R. Peres, and A. K. Geim, Science {\bf 320}, 1308 (2008).

\bibitem{graphene:OC-2} Z. Q. Li, E. A. Henriksen, Z. Jiang, Z. Hao, M. C. Martin, P. Kim, H. L. Stormer, and D. N. Basov, Nat. Phys. {\bf 4}, 532 (2008).

\bibitem{graphene:OC-3} K. F. Mak, M. Y. Sfeir, Y. Wu, C. H. Lui, J. A. Misewich, and T. F. Heinz, Phys. Rev. Lett. {\bf 101}, 196405 (2008).

\bibitem{oc-exp-1} K. Ueda, J. Fujioka, Y. Takahashi, T. Suzuki, S. Ishiwata, Y. Taguchi, and Y. Tokura, Phys. Rev. Lett. {\bf 109}, 136402 (2012).

\bibitem{oc-exp-2}  T. Timusk, J. P. Carbotte, C. C. Homes, D. N. Basov, S. G. Sharapov, Phys. Rev. B {\bf 87}, 235121 (2013).

\bibitem{oc-exp-3} R. Y. Chen, S. J. Zhang, J. A. Schneeloch, C. Zhang, Q. Li, G. D. Gu, and N. L. Wang, Phys. Rev. B {\bf 92}, 075107 (2015).

\bibitem{oc-exp-4} A. B. Sushkov, J. B. Hofmann, G. S. Jenkins, J. Ishikawa, S. Nakatsuji, S. Das Sarma, and H. D. Drew, Phys. Rev. B {\bf 92}, 241108 (2015).

\bibitem{oc-exp-5} A. Akrap, M. Hakl, S. Tchoumakov, I. Crassee, J. Kuba, M. O. Goerbig, C. C. Homes, O. Caha, J. Nov\'ak, F. Teppe, W. Desrat, S. Koohpayeh, L. Wu, N. P. Armitage, A. Nateprov, E. Arushanov, Q. D. Gibson, R. J. Cava, D. van der Marel, B. A. Piot, C. Faugeras, G. Martinez, M. Potemski, and M. Orlita, Phys. Rev. Lett. {\bf 117}, 136401 (2016).

\bibitem{oc-exp-6} D. Neubauer, J. P. Carbotte, A. A. Nateprov, A. L\"ohle, M. Dressel, and A. V. Pronin, Phys. Rev. B {\bf 93}, 121202(R) (2016).


\bibitem{TI:reivew-1}  M. Z Hasan, and C. L. Kane, Rev. Mod. Phys. {\bf 82}, 3045 (2010).

\bibitem{TI:reivew-2} X.  L.  Qi,  and  S.-C.  Zhang,  Rev.  Mod.  Phys. {\bf 83}, 1057 (2011).

\bibitem{weyl-review-1} N. P. Armitage, E. J. Mele, A. Vishwanath, arXiv:1705.01111

\bibitem{zinn-justin} J. Zinn-Justin, \emph{Quantum Field Theory and Critical Phenomena} (Oxford University Press, Oxford, UK, 2002).

\bibitem{salam1975} A. Salam and J. Strathdee, Nucl. Phys. B {\bf 90}, 203 (1975).

\bibitem{blau1991} S. K. Blau,  M. Visser and A. Wipf, Int. J. Mod. Phys. A {\bf 6}, 5409 (1991).

\bibitem{goswami-chakravarty} P. Goswami, S. Chakravarty, Phys. Rev. Lett. {\bf 107}, 196803 (2011).

\bibitem{roy-sau} B Roy, J. D. Sau, Phys. Rev. B {\bf 92}, 125141 (2015).

\bibitem{roy-goswami-sau} B. Roy, P. Goswami, J. D. Sau, Phys. Rev. B {\bf 94}, 041101 (2016).

\bibitem{juricic-balatsky} V. Juri\v ci\' c, D. S. L. Abergel, A. V. Balatsky, Phys. Rev. B {\bf 95}, 161403 (2017).


\bibitem{prokofev} I. S. Tupitsyn, and N. V. Prokof'ev, Phys. Rev. Lett. {\bf 118}, 026403 (2017).


\bibitem{vishwanath} X. Wan, A. Turner, A. Vishwanath, S. Y. Savrasov, Phys. Rev. B {\bf 83}, 205101 (2011).

\bibitem{dai-zrte} H. Weng, X. Dai, and Z. Fang, Phys. Rev. X {\bf 4}, 011002 (2014).

\bibitem{taas-2}  S-Y. Xu, I. Belopolski, N. Alidoust, M. Neupane, C. Zhang, R. Sankar, S-M. Huang, C-C. Lee, G. Chang, B. Wang, G. Bian, H. Zheng, D. S. Sanchez, F. Chou, H. Lin, S. Jia, M. Z. Hasan, Science {\bf 349}, 613 (2015).

\bibitem{nbas-1}  S-Y. Xu, N. Alidoust, I. Belopolski, C. Zhang, G. Bian, T-R. Chang, H. Zheng, V. Strokov, D. S. Sanchez, G. Chang, Z. Yuan, D. Mou, Y. Wu, L. Huang, C-C. Lee, S-M. Huang, B. Wang, A. Bansil, H-T. Jeng, T. Neupert, A. Kaminski, H. Lin, S. Jia, M. Z. Hasan, Nat. Phys. {\bf 11}, 748 (2015).

\bibitem{felser-1}  L. Yang,  Z. K. Liu,	Y. Sun,	H. Peng,	H. F. Yang,	T. Zhang,	B. Zhou,	Y. Zhang,	Y. F. Guo,	M. Rahn,	D. Prabhakaran,	Z. Hussain,	S.-K. Mo,	C. Felser,	B. Yan and Y. L. Chen, Nat. Phys. {\bf 11}, 728 (2015).

\bibitem{nbp-1}  C. Shekhar,  A. K. Nayak,	Y. Sun,	M. Schmidt,	M. Nicklas,	I. Leermakers,	U. Zeitler,	Y. Skourski,	J. Wosnitza,	Z. Liu,	Y. Chen,	W. Schnelle,	H. Borrmann,	Y. Grin,	C. Felser	and B. Yan, Nat. Phys. {\bf 11}, 645 (2015).

\bibitem{tap-1}  N. Xu, H. M. Weng, B. Q. Lv, C. E. Matt, J. Park, F. Bisti, V. N. Strocov, D. Gawryluk, E. Pomjakushina, K. Conder, N. C. Plumb, M. Radovic, G. Autès, O. V. Yazyev, Z. Fang, X. Dai, T. Qian, J. Mesot, H. Ding and M. Shi, Nat. Commun. {\bf 7}, 11006 (2016).

\bibitem{cdas}  S. Borisenko, Q. Gibson, D. Evtushinsky, V. Zabolotnyy, B. B\"uchner, and R. J. Cava, Phys. Rev. Lett. {\bf 113}, 027603 (2014).

\bibitem{nabi} Z. K. Liu, B. Zhou, Y. Zhang, Z. J. Wang, H. M. Weng, D. Prabhakaran, S.-K. Mo, Z. X. Shen, Z. Fang, X. Dai, Z. Hussain, Y. L. Chen, Science {\bf 343}, 864 (2014).

\bibitem{goswami-roy-dassarma} P. Goswami, B. Roy, S. Das Sarma, Phys. Rev. B {\bf 95}, 085120 (2017).


\bibitem{nielsen} H. B. Nielsen, and M.  Ninomiya, Phys. Lett. B {\bf 105}, 219 (1981).

\bibitem{Schmalian} In a two-dimensional Dirac system, as graphene, such crossover boundaries scale as $\Omega^\ast \sim v |n|^{1/2}$ or $T^\ast \sim (\hbar v/k_B) |n|^{1/2}$; see D. E. Sheehy, J. Schmalian, Phys. Rev. Lett. {\bf 99}, 226803 (2007).

\bibitem{Sachdev-book} S. Sachdev, \emph{Quantum Phase Transitions} (Cambridge University Press, 2nd ed., 2007).

\bibitem{hosur} P. Hosur, S. A. Parameswaran, A. Vishwanath, Phys. Rev. Lett. {\bf 108}, 046602 (2012)

\bibitem{rosenstein}  B. Rosenstein, M. Lewkowicz, Phys. Rev. B {\bf 88}, 045108 (2013).

\bibitem{roy-juricic-dassarma} B. Roy, V. Juri\v ci\' c, S. Das Sarma, Sci. Rep. {\bf 6}, 32446 (2016); B. Roy, R-J. Slager, V. Juri\v ci\' c, arXiv:1610.08973

\bibitem{Gonzalez1994} J. Gonzalez, F. Guinea, M. A. H. Vozmediano, Nucl. Phys. B {\bf 424}, 595 (1994).

\bibitem{throckmorton} R. E. Throckmorton, J. Hofmann, E. Barnes, S. Das Sarma, Phys. Rev. B {\bf 92}, 115101 (2015).

\bibitem{nandkishore} J. Maciejko, and R. Nandkishore, Phys. Rev. B {\bf 90} 035126 (2014).



\bibitem{roy-dassarma} B. Roy, S. Das Sarma, Phys. Rev. B {\bf 94}, 115137 (2016).

\bibitem{roy-goswami-juricic} B. Roy, P. Goswami, V. Juri\v ci\' c, Phys. Rev. B {\bf 95}, 201102(R) (2017).

\bibitem{Hooft} G. 't Hooft, and M. J. T. Veltman, Nucl. Phys. {\bf B 44}, 189 (1972).

\bibitem{peskin} M. E. Peskin and D. V. Schroeder, \emph{An Introduction to Quantum Field Theory} (Addison-Wesley, Reading, MA, 1995).


\bibitem{herbut-juricic-vafek-2}  V. Juri\v ci\' c, O. Vafek, I. F. Herbut, Phys. Rev. B {\bf 82}, 235402 (2010).

\bibitem{herbut-juricic-vafek-1}  I. F. Herbut, V. Juri\v ci\' c, O. Vafek, Phys. Rev. Lett.  {\bf 100}, 046403 (2008).




\bibitem{dornhaus}  R. Dornhaus, G. Nimtz, and B. Schlicht, \emph{Narrow-Gap Semicounductors}, (Springer-Verlag, 1983).

\bibitem{roy-dzero}  B. Roy, J. D. Sau, M. Dzero, V. Galitski,  Phys. Rev. B {\bf 90}, 155314 (2014).

\bibitem{katsnelson}  D. L. Boyda, V. V. Braguta, M. I. Katsnelson, M. V. Ulybyshev, Phys. Rev. B {\bf 94}, 085421 (2016).

\end{thebibliography}
\end{document}